\documentclass{amsart}
\usepackage{amsmath,amssymb}
\usepackage{pb-diagram}
\usepackage{mathpazo}
\usepackage{euler}
\usepackage{epic}
\usepackage{palatino}
\usepackage[active]{srcltx}
\numberwithin{equation}{section}
\newcommand{\M}{{\mathfrak{M}}}
\newcommand{\A}{{\mathcal{A}}}
\newcommand{\G}{{\mathcal{G}}}

\newcommand{\h}{{\mathcal{H}}}
\newcommand{\e}{{\mathcal{E}}}
\newcommand{\calA}{\mathcal{A}}

\newcommand{\calJ}{\mathcal{J}}

\newcommand{\proj}{{\mathbb{P}}}
\newcommand{\cat}{{\mathcal{C}}}
\newcommand{\Z}{{\mathbb{Z}}}
\newcommand{\R}{{\mathbb{R}}}
\newcommand{\N}{{\mathbb{N}}}
\newcommand{\C}{{\mathbb{C}}}

\newcommand{\T}{{\mathbb{T}}}

\newcommand{\sr}{/\!\!\!/}
\newcommand{\Hom}{{\operatorname{Hom}}}
\newcommand{\Ind}{{\operatorname{Ind}}}
\newcommand{\Lie}{{\operatorname{Lie}}}
\newcommand{\Det}{{\operatorname{Det}}}

\newcommand{\torus}{{\mathfrak{t}}}
\newcommand{\tr}{{\mbox{tr}}}

\def\Sr#1#2{\left.\left.#1\right\slash\hspace{-0.16cm}\right\slash\hspace{-0.08cm}#2}

\theoremstyle{plain}
        \newtheorem{theorem}{Theorem}[section]
        \newtheorem{lemma}[theorem]{Lemma}
        \newtheorem{proposition}[theorem]{Proposition}
        \newtheorem{corollary}[theorem]{Corollary}

\theoremstyle{definition}
        \newtheorem{definition}[theorem]{Definition}
        \newtheorem{remark}[theorem]{Remark}
        \newtheorem{example}[theorem]{Example}

\title{The Heisenberg group and conformal field theory}
\author{Hessel Posthuma}
\address{Korteweg-de Vries Institute for Mathematics, University of Amsterdam, P.O. Box 94248 
1090 GE Amsterdam 
The Netherlands} 
\email{H.B.Posthuma@uva.nl} 
\date{June 22, 2007. \textit{Revised}: November 25, 2009}
\begin{document}
\begin{abstract}
A mathematical construction of the conformal field theory (CFT) associated to a compact torus, also called the ``nonlinear Sigma-model'' or ``lattice-CFT'', is given. Underlying this approach to CFT is a unitary modular functor, the construction of which follows from a ``Quantization commutes with reduction''- type of theorem for unitary quantizations of the moduli spaces of holomorphic torus-bundles and actions of loop groups. This theorem in turn is a consequence of general constructions in the category of affine symplectic manifolds and their associated generalized Heisenberg groups.
\end{abstract}
\maketitle
\tableofcontents
\section*{Introduction}
The aim of this paper is to give a construction of certain conformal field theories associated to a compact abelian Lie group $T$ using the 
representation theory of the associated loop group. In physics terminology this conformal field theory is called the abelian WZW-model and is related to the $\sigma$-model on $T$, in the vertex algebra literature its chiral parts are usually referred to as the lattice-model associated to $\pi_1(T)$. It is an
abelian version of the WZW-model which describes strings moving on an arbitrary compact Lie group. 
Although in physics terminology this theory is almost ``free'', no complete mathematical account exists
in the literature. This paper fills that gap.

For this we use the mathematical axiomatisation of conformal field theory given by Graeme Segal in \cite{segal}.
This approach to conformal field theory can be paraphrased by the statement that a conformal field theory is nothing but a projective representation 
of the two-dimensional complex cobordism category. Despite the beauty and transparency of this definition, so far not many examples of this 
structure have been rigorously constructed. To the author's knowledge, two classes of examples are known to exist: first the chiral case of fermions on Riemann surfaces and its spin versions, see \cite{segal,kriz}, and secondly the so-called ``$\sigma$-model'' of a torus, cf. \cite{segal}. In both examples the partition function associated to a Riemann surface is defined in terms of elementary properties of either fermions, resp. bosons, using the fact that ``fields'' (i.e., bosons or fermions) that extend holomorphically to the surface form a maximal commutative subalgebra. The $\sigma$-model of a torus is briefly discussed in \S \ref{sigma} of this paper.

Although arguably much simpler than the non-abelian WZW-model, the abelian case is interesting enough in the sense that it is an example of a rational, but
\textit{non-chiral} conformal field theory based on a higher dimensional modular functor. To wit, the above mentioned chiral conformal field theory 
of fermions is given by a one-dimensional modular functor. On the other hand, the case of a torus stands out -in comparison with the non-abelian 
case- due to the presence of an important extra symmetry structure given by the Heisenberg group, and it is exactly this feature that we exploit in 
this paper. Indeed the moduli space of flat $T$-bundles over a closed surface is an abelian variety, and therefore its geometric quantization carries 
an irreducible representation of a Heisenberg extension of a finite group, a fact which goes back to the fundamental papers \cite{mumford66}. 
Since we need to address gluing of surfaces, we shall introduce surfaces with boundaries. On the level of moduli spaces, this leads to an infinite 
dimensional version of the theory of Abelian varieties, coined ``affine symplectic manifolds'', whose quantization corresponds to representations of 
certain associated infinite dimensional Heisenberg groups. In the infinite dimensional case, the concept of a polarization of such a variety plays a 
crucial role. 

In this approach to conformal field theory, unitarity is of the utmost importance. Indeed its construction follows from proving that a certain 
well-known modular functor is unitary. 
Although this result is probably ``known to experts'', no written account is available and the precise statement and proof is quite subtle due to the presence of the conformal anomaly.
The proof of this paper starts from a new construction of this modular functor as multiplicity spaces of certain positive energy 
representations of loop groups associated to Riemann surfaces. Factorization, in a unitary fashion, then follows from ``gluing laws'' for these 
representation, which in turn can be interpreted as a ``quantization commutes with reduction''-type of theorem. 

As remarked above, our approach differs quite a bit from the ones in literature:
geometric quantization of the finite dimensional Abelian moduli 
space and its connection with the theory of 
$\Theta$-functions is briefly discussed in \cite{atiyah}. In particular, here it is mentioned  that the 
projective flatness of the resulting quantization can 
be derived as a ``cohomological rigidity'' using the representation theory of the associated finite 
Heisenberg group, see also \cite{ramadas}. This 
idea plays a central role in the present paper. From the point of view of conformal field theory, the 
abelian, or lattice, case is briefly discussed in \cite{segal}--see 
also the ``new'' introduction to this paper--, and \cite{hk}. 
Remark that the construction of the modular functor in the present paper, namely via the moduli 
spaces of flat connections, is quite different; the connection is given in \S \ref{conformal_blocks}.
Finally, in \cite{andersen}, the relation between geometric and deformation quantization of the abelian 
moduli spaces is discussed.

\subsubsection*{\textbf{Outline of the paper}}
This paper is organized in the following way: the first section is devoted to affine symplectic manifolds, their quantization and reduction. In section 
\ref{mh} it is proved that the moduli space of flat $T$-bundles fits into this scheme, which allows us to quantize them to positive energy 
representations of $LT$, and use reduction to provide ``gluing laws''  under sewing of surfaces. This leads in section \ref{sectioncft} to the 
construction of a unitary modular functor, which in turn is used to construct the conformal field theory. 
Finally, we give a proof of the fact, stated in the introduction to \cite{segal}, that the conformal 
thus constructed is the same as the $\sigma$-model for a rational torus.
\subsubsection*{\textbf{Acknowledgement}}
The author is deeply indebted to Graeme Segal for generously sharing his ideas on the subject with 
him. He thanks Keith Hannabuss for discussions about $\Theta$-functions and pointing out reference 
\cite{mackey} to him, as well as Andre Henriques for a careful reading and suggesting several improvements. This research is financially supported by NWO.
\section{Quantization of affine symplectic manifolds}
\label{quantization}
\subsection{Affine symplectic manifolds}
\label{asm}
In this section we describe a certain category of symplectic manifolds, possibly infinite dimensional, which can be quantized in a rather 
straightforward way. This defines the appropriate framework from which we approach the moduli space of flat Abelian connections in Section 
\ref{mh}. Let us remark from the outset that all infinite dimensional manifolds in this paper will be modelled on complete nuclear topological vector 
spaces.
\begin{definition}
\label{as}
An affine symplectic manifold $(X,\omega)$ is a weakly symplectic manifold modelled on $V$ which carries a symplectic $V$-action whose isotropy 
groups $V_x,~x\in X$ are finitely generated lattices.
\end{definition}
Thus $T_xX$ can be identified with $V$ for each $x\in X$, and $V_x\cong\pi_1(X,x)$.  It is easy to see that the symplectic action of $V$ on $X$ 
furnishes $V$ with a weak linear symplectic form, also denoted $\omega$. Remark that, after choosing a basepoint $x_0\in X$, an affine 
symplectic manifold is just an (infinite dimensional) abelian group. Acting with $V$ on the basepoint, an affine symplectic manifold fits into an 
exact sequence 
\begin{equation}
\label{es-as}
0\longrightarrow \pi_1(X,x_0)\longrightarrow V\stackrel{\pi}{\longrightarrow} X\longrightarrow\pi_0(X,x_0)\longrightarrow 0.
\end{equation}
In the following, we will freely choose a basepoint $x_0$ although all constructions are actually (up to isomorphism) independent of such a choice. Notice 
that in the main examples of this paper, moduli spaces of flat bundles, there is in fact a canonical basepoint, viz. the class of the trivial bundle.

If the symplectic form $\omega$ represents an integral class in $H^2(X)$, i.e., if $\omega|_{V_{x_0}\times V_{x_0}}$ is integral, then we can find 
hermitian line bundles $(L,h)$ on $X$ with unitary connections $\nabla$ which have curvature $-\sqrt{-1}\omega$. In the language of geometric 
quantization, such a line bundle is called a prequantum line bundle. The set of isomorphism classes of prequantum line bundles forms 
a torsor for the finite dimensional torus $H^1(X,\T)$ of flat line bundles. 

Given such an $L$, we have a homomorphism $V\rightarrow H^1(X,\T)$ given by $$v\mapsto [L^*\otimes v^*L].$$ Let $V_X$ be its kernel. (Note 
that $\pi_0(V_X)\cong H^1(X,\Z)\cong\Hom(V_{x_0},\Z).)$ Define 
$$
\widetilde{V}_X:=\left\{\begin{array}{l}\mbox{connection-preserving bundle automorphisms of $(L,\nabla,h)$}\\ \mbox{which cover the action of an 
element of $V_X$ on $X$}\end{array}\right\}.
$$
\begin{proposition}
\label{hg}
The group $\widetilde{V}_X$ is a central extension $$0\rightarrow H^0(X,\T)\rightarrow\widetilde{V}_X\rightarrow V_X\rightarrow 0.$$ When $X$ is 
connected, this is the (degenerate) Heisenberg group of $(V_X,\omega|_{V_X\times V_X})$ with center $\tilde{V}_{x_0}$.
\end{proposition}
The proof of this proposition is obvious. With hindsight notice the following: 
\begin{proposition}
For $X$ connected, there is an equivalence of categories:
$$
\left\{\begin{array}{c}\mbox{{\rm Prequantum line bundles}}\\ \mbox{{\rm over} $(X,\omega)$.}\end{array}\right\}\cong\left\{\begin{array}{c}\mbox{{\rm 
Splittings of the extension} $\tilde{V}_{x_0}$}\\ \mbox{{\rm induced from the Heisenberg group} $\tilde{V}$.}\end{array}\right\}
$$
\end{proposition}
Remark that $V_{x_0}$ is an isotropic lattice in $(V,\omega)$. The functor establishing the equivalence above sends a splitting $\chi$, viewed as a 
character $\chi:\tilde{V}_{x_0}\rightarrow\T$, to the line bundle 
$$
L_\chi:=\tilde{V}\times_{\tilde{V}_{x_0}}\C_\chi.
$$ 
Conversely, the fiber over $x_0$ of a prequantum line bundle carries a unitary representation of $\tilde{V}_{x_0}$, i.e., determines a character $\chi:
\tilde{V}_{x_0}\rightarrow\T$. Also notice that the set of splittings of $\tilde{V}_{x_0}$ forms a torsor over the Pontryagin dual 
$\hat{V}_{x_0}:=\Hom(V_{x_0},\T)$, which corresponds to the action of the torus of flat line bundles using the isomorphism $\Hom(V_{x_0},\T)\cong 
H^1(X,\T)$. 
\begin{definition}
\label{pol-afs}
A polarization of an affine symplectic manifold $X$ modelled on $V$ means a polarization of $V$.
\end{definition}
A polarization of a symplectic vector space is a positive compatible complex structure, as explained in the appendix, cf. Definition 
\ref{pol-vs}. A polarized affine symplectic manifold is simply an affine symplectic manifold together with the choice of a polarization. Notice that 
when the manifold is a finite dimensional symplectic torus, this notion coincides with that of a polarized Abelian variety.
\subsection{Quantization}
\label{squantization}
In this section we shall construct quantizations of affine symplectic manifolds. Since this construction
works for each connected component separately, we will assume that our affine symplectic manifold $X$ is connected.
Assume for a moment that it is finite dimensional. The way to proceed in geometric quantization is to define, for 
each polarization, the Hilbert space $H^0_{L^2}(X;L):=L^2_{hol}(X,L)$ of holomorphic sections of a prequantum line bundle $(L,h)$ with inner 
product given by $$\left<s_1,s_2\right>:=\int_X h(s_1,s_2)\frac{\omega^n}{n!},$$ where $\dim X=2n$. Going over to the infinite dimensional case, 
one immediately realizes that the Liouville measure doesn't exist as there are no nontrivial invariant measures on an infinite dimensional vector 
space. However, below we will show that for affine symplectic manifolds, the combination ``$h(-,-)\omega^n/n!$'' does make sense as a line bundle 
valued measure (cf.\ \cite{pickrell}, and see below). The rationale behind this is that for an infinite dimensional vector space, this combination 
formally combines into a Gaussian measure which can be rigorously constructed in infinite dimensions. 

Let $(X,L)$ be a polarized affine symplectic manifold modelled on $V$. First of all, a thickening of $X$ is a smooth manifold $X^*$ modelled on 
$V^*$, the dual, equipped with a continuous dense embedding $X\hookrightarrow X^*$. More precisely, the polarization $J$ puts $V$ and $V^*$ in a triple 
$V\subseteq H_J\subseteq V^*$, where $H_J$ is an intermediate Hilbert space, the completion of $V$ in the inner product defined by $\omega$ 
and $J$, cf.\ Appendix \ref{heisenberg}. Therefore, we aim for a ``rigged manifold'' $$X\hookrightarrow X_H\hookrightarrow X^*,$$ where the 
intermediate space $X_H$ is a Hilbert manifold modelled on $H_J$. We will prove below that there exists a canonical thickening of any affine 
symplectic manifold and that the line bundle $L$ extends to $X^*$. Then, a measure $\mu^L$ on $X^*$ with values in $L$ consists of a continuous 
map $$\mu^L:\Gamma(X^*,L)\times\Gamma(X^*,L)\rightarrow \left\{\mbox{complex Borel measures on}~ X^*\right\},$$ written $(s_1,s_2)\mapsto 
\mu^L_{s_1,s_2}$, or $h(s_1,s_2)d\mu$, satisfying the following properties: 
\begin{itemize}
\item  for all $f_1,f_2\in C(X)$ we have $$\mu^L_{f_1s_1,f_2s_2}=\bar{f}_1f_2\mu_{s_1,s_2},$$ i.e., the map is sesquilinear as as a map of $C(X)$-modules,
\item the measure is positive in the sense that $$\mu^L_{s,s}\geq 0,$$ for all sections $s\in\Gamma(X^*,L)$, with strict inequality for $s\neq 0$,
\item for $s_i\in\Gamma(X^*,L),~i=1,\ldots,4$, the measure $\mu^L_{s_1,s_2}$, restricted to to the set $\{x\in X^*,~h_L(s_3(x),s_4(x))\neq 0\}$, is absolutely continuous with respect to $\mu^L_{s_3,s_4}$ with Radon--Nikodym derivative equal to $$\frac{d\mu^L_{s_1,s_2}}{d\mu^L_{s_3,s_4}}=\frac{h(s_1,s_2)}{h(s_3,s_4)}.$$
\end{itemize}
\begin{proposition}
\label{lvm}
An affine symplectic manifold $X$ has a canonical thickening $X^*$ such that any prequantum line bundle $L$ extends to $X^*$, and a 
polarization $J$ of $X$ defines a unique $\widetilde{V}_{X}$-invariant measure $\mu^L_J$ on $X^*$ with values in $L$. 
\end{proposition}
\begin{proof}
First consider the statement for a symplectic vector space $(V,\omega)$. The, up to isomorphism, unique prequantum line bundle may be realized 
as the trivial bundle $L=V\times \C$ equipped with the Hermitian metric $$h((v,z_1),(v,z_2))=e^{-\left<v,v\right>/2}\bar{z}_1z_2,$$ where 
$\left<~,~\right>$ is the metric defined by the polarization $J$, cf.\ \eqref{hermmet}. Indeed one computes $-\partial\bar{\partial}\log h=-\sqrt{-1}\omega$. 
Therefore, the action of $V$ by translations lifts to connection preserving transformations of $V\times\C$ by the formula 
\begin{equation}
\label{al}
v\cdot(w,z)=(v+w,e^{\left<v,v\right>/4+\left<v,w\right>/2}z).
\end{equation} 
This defines an action of the Heisenberg group $\tilde{V}$ associated to the symplectic vector space $(V,\omega)$; 
this gives an explicit realization of Proposition \ref{hg} above.

As described in Appendix \ref{lc}, a compatible positive complex structure $J$ on $(V,\omega)$ defines a family of Gaussian measures 
$\mu^t_J,~t>0,$ on $V^*$, the dual of $V$, together with a dense embedding $V\hookrightarrow V^*$.
 Put $\mu_J:=\mu^1_J$, extend the trivial bundle $L$ to $V^*$, and define, for a continuous function 
$F:V^*\rightarrow\C$; $$d\mu^L_{F_1,F_2}:=\bar{F}_1F_2d\mu_J.$$ It is obvious that when $F$ is nonzero, this defines a positive Borel measure. 
Acting by $\tilde{V}$ as induced from \eqref{al}, this measure is invariant since the exponential factors in $\bar{F} F$ cancel against the contribution 
of the Cameron--Martin formula \eqref{cm}. We thus observe that in the linear case, the Proposition is merely a reformulation of the existence and 
properties of the Gaussian measure, which indeed is uniquely determined by the polarization $J$.

When $V$ is finite dimensional, the Gaussian measure thus defined can of course be written as the product 
$$d\mu_J(v)=e^{-\left<v,v\right>/2}\frac{\omega^n}{n!},$$ where $\omega^n/n!$ is the Liouville measure, i.e., a Haar measure on $V$. In fact this is 
the expression for any finite dimensional affine symplectic manifold: the $L$-valued measure is defined as 
$$\mu_{s_1,s_2}^L:=h(s_1,s_2)\frac{\omega^n}{n!},$$ where $h$ is the Hermitian metric on $L$. 
Finally, in the general case, notice we can choose a decomposition $X=W\times Y$, with $W$ a symplectic vector space and $Y$ a finite dimensional affine symplectic manifold. This reduces the 
general statement to the two special cases considered above, and the resulting measure is independent of the decomposition. By construction, this line bundle valued measure is invariant.  
\end{proof}
With the help of this measure one can define the quantization of $X$ to be the $L^2$-space of holomorphic sections of $L$ on $X^*$ with the 
inner-product given by 
\begin{equation}
\label{inprgq}
\left<s_1,s_2\right>:=\int_{X^*}h(s_1,s_2)d\mu.
\end{equation}
We denote this Hilbert space by $\h_X:=L^2_{\mbox{\tiny hol}}(X,L)$. 
 When we want to stress the dependence on the line bundle $L$ in the notation, we write $\h_{X,L}$ instead of $\h_X$.
\begin{theorem}
\label{irrheis}
$\h_{X,L}$ carries an irreducible unitary representation of $\widetilde{V}_{X}$, with the center acting by the character corresponding to $L$.
\end{theorem}
\begin{proof}
Since the measure $\mu^L_J$ is $\widetilde{V}_{X}$-invariant, $\h_X$ carries a representation of $\widetilde{V}_{X}$ which is unitary for the inner 
product defined by \eqref{inprgq}. The statement about the action of the center follows immediately from the definitions. We need to show that the 
representation is irreducible. Bu this follows easily from the decomposition $X\cong W\times Y$ as in 
the proof of Proposition \ref{lvm}: this reduces the statement to the Examples \ref{example} $i)$ and $ii)$ below. 
\end{proof}
\begin{corollary}
\label{ics}
The quantization $\h_X$ is, up to isomorphism,  independent of the specific complex structure in the polarization class.
\end{corollary}
\begin{proof}
This follows immediately from Theorem \ref{irgh}, which states that, for a given polarization class, 
the Heisenberg extension of $V_{X}$ has only one irreducible representation for a fixed character of the center.
\end{proof}
\begin{proposition}
\label{indaf}
For $X$ connected, there is an isomorphism of $\tilde{V}$-representations 
\[
\Ind_{\tilde{V}_X}^{\tilde{V}}(\h_X)\cong\h_V.
\]
\end{proposition}
\begin{proof}
Let us first recall the construction of the induced representation in this case. Consider the inclusion $V_X\hookrightarrow V$ and notice that there is 
an exact sequence $$0\rightarrow V_X\rightarrow V\rightarrow X^d\rightarrow 0,$$ where $X^d\cong\Hom(V_{x_0},\T)$ is a finite dimensional 
torus. Therefore we have $$\Ind_{\tilde{V}_X}^{\tilde{V}}(\h_X):=L^2\left(X^d,\h_X\right)$$ with the inner product defined with respect to the Haar 
measure on $X^d$. But since $X^d$ is Pontryagin dual to $V_{x_0}$, the Hilbert space above is exactly the spectral decomposition of $\h_V$ with 
respect to the abelian subgroup $V_{x_0}\subset \tilde{V}$. 
\end{proof}
Since the representation of $\tilde{V}$ on $\h_V$ is irreducible, this gives another proof of irreducibility of the representation of $\tilde{V}_X$ on $\h_X$ by Mackey's theorem. Of course, a similar result holds for non-connected $X$.
\begin{example}
\label{example}
Let us give some examples of this quantization:
\begin{enumerate}
\item[$i)$] When $X=V$ is a symplectic vector space, the quantization $\h_V$ is simply the standard irreducible representation of the Heisenberg group $\tilde{V}$ associated to the given polarization \cite{bsz}. Instead of the construction above with the Gaussian measure, one can construct the representation directly on the symmetric Hilbert space $\operatorname{Sym}(W)$, where $V_\C=W\oplus\bar{W}$ is the decomposition into $+\sqrt{-1}$ and $-\sqrt{-1}$ eigenspaces of $J$, cf.\ \cite{segalcmp}.
\item[$ii)$] In case of a finite dimensional polarized abelian variety $(X,L)$, the prescription above produces the Hilbert space $H^0(X,L)$ equipped with the Liouville inner product. Of course, this is just the space of $\Theta$-functions, which, by the Riemann--Roch theorem, has dimension equal to the symplectic volume of $X$. In this case, $\tilde{V}_X$ is a finite Heisenberg group. That $H^0(X,L)$ is the unique irreducible representation of this group is well known, see e.g. \cite[\S 1.3]{mumford,po}. The torus $H^1(X,\T)$ parameterising inequivalent quantizations of $X$ is in this context usually referred to as the \textit{dual torus} $X^d$.
\end{enumerate}
\end{example}
\subsection{The universal family} Next, we will introduce, for each complex structure in the polarization class, a canonical rigging of the Hilbert space $\h_X$. Since this Hilbert space consists of square integrable sections of an extension of the line bundle $L$ to $X^*$, one obtains, by restriction to $X$, a canonical map $\h_X\rightarrow \hat{E}_X$, where $\hat{E}_X:=\Gamma_{hol}(X,L)$. This map is injective by the fact that the inclusion $X\hookrightarrow X^*$ is dense, i.e., holomorphic sections of $L$ on $X^*$ are uniquely determined by their restriction to $X$. When equipped with the topology of uniform convergence on compact subsets, $\hat{E}_X$ is a complete, locally convex topological vector space, and the map $\h_X\hookrightarrow\hat{E}_X$ is a dense continuous inclusion. Let $\check{E}_X$ be the continuous anti-dual of $\hat{E}_X$, i.e., the complex conjugate space of all linear continuous functionals on $\hat{E}_X$. Restricting such a functional to $\h_X$ defines, by Riesz' theorem, a vector in $\h_X$. This defines a dense embedding of $\check{E}_X$ into $\h_X$ and gives a rigging of this Hilbert space;
\begin{equation}
\label{rigging}
\check{E}_X\hookrightarrow\h_X\hookrightarrow\hat{E}_X,
\end{equation}
where each of the inclusions is dense and continuous. For $l\in L\backslash\{0\}$, evaluation at $l$ defines an element $ev_l\in\check{\Gamma}_X$, because the sections are holomorphic. By the inclusions above, this gives maps as in the following diagram:
$$\begin{diagram} \node{\overline{L}\backslash\{0\}}\arrow{e,t}{}\arrow{s}\node{\overline{\h}_X\backslash\{0\}}\arrow{s}\\\node{X}\arrow{e}\node{\mathbb{P}\h_X}\end{diagram}$$
Notice that the right hand side is just the universal holomorphic line bundle over the projective Hilbert space. The inner-product in $\h_X$ induces a natural hermitian metric on this bundle whose curvature is given by the Fubini--Study form on $\mathbb{P}(\h_X)$. Now we have:
\begin{proposition}
\label{cs}
The map $X\rightarrow\proj(\h_X)$ is an embedding of K\"{a}hler manifolds, i.e., the pull-back of the Fubini--Study form on $\mathbb{P}(\h_X)$ equals $\omega$. Consequently, the pull-back of the universal bundle to $X$ is isomorphic, via the upper map in the diagram, to the prequantum line bundle $L$.
\end{proposition}
\begin{proof}
It is easy to check that the map $X\rightarrow\mathbb{P}(\h_X)$ is a holomorphic embedding. To see that it is in fact K\"ahler, consider $\mathbb{P}(\h_X)$ as a 
K\"ahler manifold with symplectic form given by the Fubini--Study metric. As stated above, the dual of the universal bundle given by $\h_X\backslash\{0\}$ gives 
a pre-quantization of $\mathbb{P}(\h_X)$. Because $V_X$, by its embedding in $PU(\h_X)$, acts on $\mathbb{P}(\h_X)$ by K\"ahler isometries, this 
prequantum line bundle determines a central extension of $V_X$ uniquely determined by the Fubini--Study form. But this central extension must be equal to the 
Heisenberg extension of $V_X$ defined by the symplectic form $\omega$ on $X$ since this is the group that actually acts by unitary transformations on $\h_X$.

Next consider the pull-back of the universal bundle to $X$. By the argument above, its curvature equals $\omega$. 
Since the projective representation of $V_X$ 
on $\h_X$ corresponds to the same character of $V_{x_0}$ as the one that determines the line bundle $L$ over $X$, cf.\ Proposition \ref{irrheis}, the two line 
bundles are isomorphic.
\end{proof}
This proposition gives a complete characterization of the Hilbert space $\h_{X,L}$ quantizing 
an affine symplectic manifold $(X,L)$ with a prequantum line bundle for a given polarization as 
in the theorem below. In the following, let $\calJ(X)$ be the space of polarizations of $X$ as in 
Definition \ref{pol-afs}. Recall that this is just the Siegel upper half space $\mathcal{J}(V)$
of the symplectic vector space $V$ modelling $X$, cf.\ \S \ref{ts}.
\begin{theorem}
\label{qbundle}
For a fixed pair $(X,L)$ consisting of an affine symplectic manifold $X$ with a prequantum line bundle 
$L$, there exists, for each polarization $J\in\calJ(X)$, a unique Hilbert space $\h_{X,L}$ satisfying:
\begin{itemize}
\item $\h_{X,L}$ carries an irreducible representation of the Heisenberg group $\tilde{V}_X$ with 
central character determined by $L$,
\item $\h_{X,L}$ comes equipped with an $\tilde{V}_X$-equivariant isometric map $L\to\h_{X,L}$.
\end{itemize}
When $J$ varies, $\h_{X,L}$ forms a projectively flat bundle of Hilbert spaces over $\calJ(X)$.
\end{theorem}
\begin{proof}
The construction of the previous section, together with Proposition \ref{cs}, prove existence of the
Hilbert space. By Corollary \ref{ics}, $\h_{X,L}$ is unique up to isomorphism, and by Schur's lemma,
this isomorphism is canonical up to a scalar. However, this scalar is precisely fixed by requirement 
that the map $L\to\h_{X,L}$ is an isometry. This proves uniqueness of the Hilbert space. 
As the polarization varies, again Corollary \ref{ics} in combination with Schur's lemma show
that the bundle $\proj\h_{X,L}$ is canonically flat.
\end{proof}
\begin{remark}
When $X$ is compact, i.e., an abelian variety, the quantization is finite dimensional and the theorem 
above implies that the connection associated to the holomorphic hermitian vector bundle $\h_X$ is 
projectively flat. 
This is the connection described in \cite{welters} leading to the heat equation satisfied by 
$\Theta$-functions.
\end{remark}
Let Aut$(X)\subseteq{\rm Sp}(V)$ be defined as $${\rm Aut}(X):=\{\mbox{Symplectic automorphisms 
of $(V_{x_0},\omega)$}\}.$$ When $X$, and therefore $V$, is polarized, define Aut$_{res}(X)={\rm 
Aut}(X)\cap {\rm Sp}_{res}(V)$, where ${\rm Sp}_{res}(V)$ is the restricted symplectic group, cf.\ \S 
\ref{ts}. Since elements of ${\rm Aut}(X)$ need not preserve the complex structure, this group a priori 
does not  
act on the quantization $\h_X$. However, elements of Aut$_{res}(X)$ map one point of $\mathcal{J}
(V)$ to another and therefore Proposition \ref{qbundle} gives:
\begin{corollary}
\label{aut}
$\h_X$ carries a projective unitary representation of {\rm Aut}$_{res}(X)$. It extends the unitary representation of $\widetilde{V}_{X}$ to the semi-direct product $\widetilde{\mbox{\rm Aut}}_{res}(X)\ltimes\widetilde{V}_{X}$
\end{corollary}
\subsection{Symplectic reduction}
\label{csr}
In this section we will study the reduction theory of affine symplectic manifolds with respect to (part of) its affine symmetries. Let $X$ be an affine symplectic manifold modelled on $V$, which is assumed to be connected. By definition, the vector space $V$ acts symplectically on $X$. Restricted to $V_X$, this action is even Hamiltonian with moment map $J:X\rightarrow \Lie(V^*_X)$ given by 
\begin{equation}
\label{moment}
J(x)=-\omega(v_x,-)|_{\Lie(V_0^*)},
\end{equation}
where $v_x\in V$ is a pre-image of $x$ under $\pi$, cf. the exact sequence \eqref{es-as}. This may be seen to be independent of the choice of the 
lift, i.e., up to the action of $V_{x_0}$, by realizing that $V_X\subseteq V$ is the projection of the commutant of $\tilde{V}_{x_0}$ in $\tilde{V}$.  
Notice that the moment map is only \textit{affine}-equivariant, with respect to the cocycle defined by the symplectic form $\omega$. This is exactly 
the cocycle that defines the Heisenberg extension of $V_X\subseteq V$. A closed abelian subgroup $A\subseteq V_X$ is said to be isotropic if the 
induced central extension $\tilde{A}$ is abelian. As such, the extension must be trivial, but not canonically, i.e., depends on the choice of a splitting 
$\chi:A\rightarrow\T$ satisfying $$\chi(a_1a_2)=\chi(a_1)\chi(a_2)e^{\sqrt{-1}\pi\omega(a_1,a_2)}.$$ Additively, this is given by a map $\psi:A\rightarrow\R$ 
satisfying $\psi(a_1+a_2)=\psi(a_1)+\psi(a_2)+\omega(a_1,a_2)$. Therefore, $J_A:=J+\psi$ defines a moment map which is $A$-equivariant and 
one can consider the symplectically reduced space $$X_{red}=\Sr{X}{A}:=J_A^{-1}(0)/A.$$
\begin{proposition}
\label{sr}
Let $(A,\chi)$ be an isotropic subgroup of $V_X$ with a choice of splitting. When $A$ acts freely on $X$, $X_{red}$ is an affine symplectic manifold modelled on 
\[
V_{red}={\rm Lie}(A)^\circ/{\rm Lie}(A),
\] 
where ${\rm Lie}(A)^\circ$ denotes the symplectic complement of $\Lie(A)\subseteq V$ with respect to $\omega$. A polarization of $X$ induces a polarization of $X_{red}$, and a prequantum line bundle $L$ on $X$ induces a prequantum line bundle $L_\chi$ on $X_{red}$. 
\end{proposition}
\begin{proof}
First notice that since $A$ is assumed to act freely, one has in fact $A\subset V_X$, and $A$ fits into a short exact sequence of abelian groups 
\[
0\rightarrow\Lie(A)\rightarrow A\rightarrow\pi_0(A)\rightarrow 0.
\] 
It is easy to see that the splitting $\psi$ must be zero on $\Lie(A)$, i.e., $\Lie(A)\subset V$ is an isotropic subspace in the sense that $\omega|_{\Lie(A)}=0$. Consider the zero locus of the moment map, $J_A^{-1}(0)\subseteq X$. It follows at once from the formula \eqref{moment} for the moment map that the symplectic complement $\Lie(A)^\circ$ is the maximal subspace of $V$ acting transitively on $J_A^{-1}(0)$. It therefore follows that $X_{red}$ carries a transitive affine action of $V_{red}$. Since $\pi_1(X)$ and $\pi_0(A)$ are assumed to be finitely generated lattices, the isotropy groups of this action will be finitely generated as well. This proves that $X\sr A$ is an affine symplectic manifold modelled on $V_{red}$ in the sense of Definition \ref{as}.
Let $L$ be a prequantum line bundle over $X$. It is easy to check that $$L_\chi:=L|_{J_A^{-1}(0)}\times_{\tilde{A}}\C_\chi$$ defines a prequantum line bundle over $X\sr A$. Finally, the fact that a polarization $J$ of $V$ induces a polarization of $\Lie(A)^\circ/\Lie(A)$, follows from the canonical isomorphism $V\sr\Lie(A)\cong V/\Lie(A)_\C$; the right hand side has a canonical complex structure compatible with the symplectic form.
\end{proof}
\begin{remark}
Notice that the first two homotopy groups of $X_{red}$ fit into an exact sequence of abelian groups 
\begin{equation}
\label{esfg}
0\rightarrow\pi_1(X)\rightarrow\pi_1(X_{red})\rightarrow\pi_0(A)\rightarrow\pi_0(X)\rightarrow\pi_0(X_{red})\rightarrow 0.
\end{equation}
Of course, the higher homotopy groups are trivial.
\end{remark}
\begin{proposition}
\label{hgrs}
The Heisenberg groups associated to affine symplectic manifolds are related under reduction by 
\[
\tilde{V}_{X_{red}}\left\slash\widetilde{\pi_1(X_{red})}\right.\cong \left.\tilde{A}^\perp\right\slash A,
\]
where the commutant $\tilde{A}^\perp$ is taken in the Heisenberg group $\widetilde{V}_{X}\left\slash\widetilde{\pi_1(X)}\right.$
\end{proposition}
\begin{proof}
The proof is straightforward, we omit the details. 
\end{proof}
\subsection{Induced representations}
\label{indrep}
In this section we discuss the quantized version of the reduction procedure of the previous section,
so that we can have that ``quantization commutes with reduction''. The upshot will be that such a procedure is best given in terms of induced representations. This is best motivated by the classical example of $\Theta$-functions. Let us start by the following remark: suppose that $\h$ is a unitary representation of 
a group $G$ which is discretely reducible, with $\h_i,~i\in I$ being the irreducible representations that can occur in its decomposition. Then the space 
$\Hom_G(\h_i;\h)$ of intertwiners carries a canonical inner product given by 
$$
\left<\psi_1,\psi_2\right>:=\psi_1^*\psi_2\in\Hom_{G}(\h_i;\h_i)=\C,
$$
 for $\psi_1,\psi_2\in \Hom_{G}(\h_i;\h).$ With this inner product, the canonical map 
\begin{equation}
\label{can_iso}
\bigoplus_{i\in I}\Hom_G(\h_i;\h)\otimes\h_i\stackrel{\cong}{\longrightarrow}\h
\end{equation}
defined by $\psi_i\otimes v_i\mapsto\psi_i(v_i)\in\h$, for $\psi_i\in \Hom_G(\h_i;\h)$ and $v_i\in\h_i$, is a unitary isomorphism providing the decomposition of 
$\h$ into irreducibles.
\begin{theorem}
\label{theta}
Let $(X,L_\chi)$ be a polarized abelian variety with $X=V/\Lambda$ where $\Lambda$ is a full isotropic lattice in a finite dimensional complex symplectic vector 
space $(V,\omega)$, with a choice of splitting $\chi$. Then the associated space of Theta-functions is given by 
$$
H^0(X,L_\chi)\cong\Hom_{\tilde{V}}\left(\Ind_{\tilde{\Lambda}}^{\tilde{V}}(\C_\chi),\h_V\right).
$$
\end{theorem}
\begin{proof}
By Proposition \ref{irrheis}, the left hand 
side carries an irreducible representation of the Heisenberg group $\tilde{\Lambda}^\circ$, with the 
center $\tilde{\Lambda}$ acting via $\chi$. Choose a Lagrangian lattice $L$ intermediate between 
$\Lambda$ and $\Lambda^\circ$ together with an extension $\chi_L$ of $\chi$; we have 
$\Lambda\subseteq L\subseteq\Lambda^\circ$, and $L$ is isotropic and maximal. Functoriality of 
induction then gives a natural isomorphism $$\Ind_{\tilde{\Lambda}}^{\tilde{V}}\cong\Ind_{\tilde{L}}^{\tilde{V}}\circ\Ind_{\tilde{\Lambda}}^{\tilde{L}}.
$$ Applied to the representation $\C_\chi$, we first compute $$\Ind_{\tilde{\Lambda}}^{\tilde{L}}
(\C_\chi)=L^2(L/\Lambda;L_\chi),$$ but this representation of $\tilde{L}$ has a canonical extension 
to $\tilde{\Lambda}_{op}^\circ$, viz. the ``Schr\"odinger representation'' associated to $L$. This is the 
irreducible representation of the Heisenberg extension $\tilde{\Lambda}^\circ_{op}$ with the center 
acting via $\chi$, i.e., precisely isomorphic to $\h^*_{X,\chi}$. With this we compute
\begin{displaymath}
\begin{split}
\Ind_{\tilde{\Lambda}}^{\tilde{V}}(\C_\chi)&\cong \Ind_{\tilde{L}}^{\tilde{V}}\left(\h^*_{X,\chi}\right)\\&=L^2(V/L;\tilde{V}\times_{\tilde{L}}\h^*_{X,\chi})\\&\cong L^2(V/L;L_{\chi_L})\otimes\h^*_{X,\chi},
\end{split}
\end{displaymath}
where in the last line we have used the splitting $\chi_L$ to trivialize the bundle of Hilbert spaces formed by $\h^*_{X,\chi}$. Now, the first factor in the tensor product is simply the representation of $\tilde{V}$ induced from $\C_{\chi_L}$, which is irreducible by Mackey's theorem and therefore killed by taking $\Hom_{\tilde{V}}(-,\h_{\tilde{V}})$. The result now follows.
\end{proof}
\begin{remark}
The subspace $\tilde{\Lambda}$-invariants in $\h_V$ is trivial, i.e., equal to $\{0\}$. This can be seen as follows: by a similar line of reasoning, this time using the 
chain of inclusions $L\subseteq\Lambda^\circ\subset V$, one finds 
\begin{displaymath}
\begin{split}
\h_V&\cong \Ind_{\tilde{\Lambda}^\circ}^{\tilde{V}}\circ\Ind_{\tilde{L}}^{\tilde{\Lambda}^\circ}(\C_\chi)\\ &\cong 
\Ind_{\tilde{\Lambda}^\circ}^{\tilde{V}}\left(\h_{\tilde{\Lambda}^\circ,\chi}\right)\\ &\cong L^2\left(X^d,\h_{X,\chi}\right).
\end{split}
\end{displaymath}
This is, of course, nothing but the spectral decomposition of $\h_V$ under the representation of the abelian subgroup $\Lambda$, acting via the splitting $\chi$. 
It clearly show that the ``reduced space'' $\h_X$ is not a closed subspace of invariants of any kind of $\h_V$. 

The fact that the Hilbert space has no nontrivial invariants, is 
usually circumvented, cf.\ \cite{mumford}, by taking invariants under $\Lambda$ in a certain distributional completion of $\h_V$. The main advantage of the point of view expressed in 
the theorem is its manifest unitarity: it only refers to the Hilbert spaces involved in the quantization. As such, one easily shows that the isomorphism is completely 
natural, i.e., as projectively flat Hilbert bundles over $\mathcal{J}(V)$. This point of view on $\Theta$-functions, i.e., as intertwiners between certain induced 
representations, originated in \cite{mackey}.
\end{remark}
The general idea should now be clear: for an isotropic subgroup $A$ acting on an affine symplectic manifold $X$,  instead of considering the (empty) subspace of $A$-invariants in $\h_X$, we induce the trivial representation given by a splitting up to a Heisenberg subgroup of $V_X$ that contains $A$, and consider the space of intertwiners. Instead of developing this in full generality, let us consider the following case: let $X_1$ and $X_2$ be two polarized affine symplectic manifolds modelled on $V_1$ and $V_2$, and suppose $A\subseteq V_{X_i}$ is such that the restriction of $\tilde{V}_{X_i}$ defines a generalized Heisenberg extension $\tilde{A}$ of $A$. In this case $A$ acts as an isotropic subgroup on the product $X_1\times\overline{X}_2$ with a canonical splitting. Assume furthermore that the center $Z(A)$ of $\tilde{A}$ is finite and that the two
inclusions $\Lie(A)\hookrightarrow V_i,~i=1,2$ induce the same polarization class on $A$. 
Let $L_1$ and $L_2$ be prequantum line bundles on $X_1$ and $X_2$.
\begin{theorem}
\label{qcras}
In this case, there is an isomorphism of $\tilde{V}_{X_{red}}$-representations
\[
\h_{X_{red},L_{red}}\cong\bigoplus_{\chi\in\widehat{Z(A)}}\Hom_{\tilde{A}}\left(\h_\chi,\h_{X_1,L_1}\right)
\otimes
\Hom_{\widetilde{A}_{op}}\left(\h^*_\chi,\h^*_{X_2,L_2}\right)
\]
\end{theorem}
\begin{proof}
Clearly, the right hand side carries a unitary representation of the generalized Heisenberg group
$
\tilde{A}_1^\perp\times_{Z(A)}\tilde{A}_2^\perp,
$
where $\tilde{A}_i^\perp,~i=1,2$ is the commutant of $\tilde{A}$ in $\tilde{V}_{X_i}$, by composition with intertwiners. Decomposing the representations $\h_{X_i,L_i}$ under the action of $\tilde{A}$ as
\[
\h_{X_i,L_i}=\bigoplus_{\chi\in\widehat{Z(A)}}\Hom_{\tilde{A}}\left(\h_\chi,\h_{X_i,L_i}\right)\otimes\h_\chi,
\]
cf.\ \eqref{can_iso}, it follows that this representation is irreducible. 
But it follows from Proposition \ref{hgrs} that this Heisenberg group is exactly $\tilde{V}_{X_{red}}$.
\end{proof}
\begin{remark}
The connection with Theorem \ref{theta} is given by the fact that
\[
{\rm Ind}_{A}^{\tilde{A}\times\tilde{A}_{op}}(\C)=\bigoplus_{\chi\in\hat{Z}_A}\h_\chi\otimes\h_\chi^*.
\]
\end{remark} 
\section{Quantization of the moduli space of holomorphic torus-bundles}
\label{mh}
\subsection{Notation} Let us briefly introduce some notation in relation to abelian gauge theory. 
Let $T$ be a compact abelian Lie group with Lie algebra $\torus$. 
For any compact manifold $M$, possibly with boundary, let $\calA(M)$ be the space of smooth connections 
on the trivial principal $T$-bundle over $M$. This is an infinite dimensional affine Fr\'echet space modelled
 on $\Omega^1(M,\torus)$, the space of $1$-forms with values in $\torus$. The gauge group 
$T(M):=C^\infty(M,T)$ is defined to be the Fr\'echet Lie group of smooth maps from $M$ into $T$, 
with Lie algebra given by $\torus(M):=\Omega^0(M,\torus)$. Writing $T=\torus/\Lambda$, 
where $\Lambda:=\Hom(\T,T)$ is the integral lattice, we have a short exact sequence 
$$
0\rightarrow\underline{\Lambda}\rightarrow\underline{\torus}\rightarrow\underline{T}\rightarrow 0,
$$
 of sheaves of smooth functions. The long exact sequence in cohomology then gives 
$$
0\rightarrow H^0(M,\Lambda)\rightarrow\torus(M)\rightarrow T(M)\rightarrow H^1(M,\Lambda)\rightarrow 0,
$$ 
from which we read off that $$\pi_0(T(M))= H^1(M,\Lambda),\hspace{0.5cm}\pi_1(T(M))= H^0(M,\Lambda).$$
The gauge group $T(M)$ naturally acts on $\calA(M)$ by the affine transformations 
\begin{equation}
\label{gauge}
\varphi\cdot A=A-d\varphi\varphi^{-1},
\end{equation}
where $A\in\calA(M)$ and $\varphi\in T(M)$. This action is generated by the fundamental vector fields $-d\xi$, where $\xi\in\Omega^0(M,\torus)$. Below, we will use these objects only in the cases that $\dim(M)=1$ or $2$.
\subsection{Positive energy representations of $LT$}
\label{iper}
In this section we will discuss the notion of a positive energy representation of the gauge group $T(S)$ associated to a compact oriented $1$-manifold $S$. When $S$ is connected, it is of course diffeomorphic to $S^1$, and the group $T(S^1)$, called the loop group of $T$, is denoted by $LT$. Let Rot$(S^1)$ be the group of rotations of the circle, i.e., the group $S^1$ acting on itself. Recall the definition of a positive energy representation \cite[\S 9]{ps}:
\begin{definition}
\label{per}
A positive energy representation of $LT$ is a projective unitary representation on a Hilbert space $\h$, given by a strongly continuous homomorphism 
$\pi:LT\rightarrow PU(\h)$, which has an extension to the semi-direct product ${\rm Rot}(S^1)\ltimes LT$ such that the action of ${\rm Rot}(S^1)\cong \T$ can be 
lifted to an action by non-negative characters.
\end{definition}
By the positive energy condition, the projective representations above are actually 
true representations of a \textit{smooth} central extension
\begin{equation}
\label{ce}
1\rightarrow\T\rightarrow \widetilde{LT}\rightarrow LT\rightarrow 1.
\end{equation}
These extensions are classified as follows: since $LT$ is abelian, any such extension is
topologically trivial and, up to isomorphism, determined by the commutator map $s:LT\times 
LT\rightarrow\T$, cf.\ \S \ref{def-hg}. Taking the fundamental group, this defines a quadratic form
$q:\Lambda\times\Lambda\rightarrow \Z$ on $\pi_1(T)$, which classifies the extension up to isomorphism.

We will now assume, cf.\ Remark \ref{level-spin} below, that this quadratic form $q$ turns the lattice
into an even one. Then the corresponding central extension can be described as follows: 
extend $q$ to a an inner product $\left<~,~\right>:\torus\times\torus\rightarrow\R$ on the Lie 
algebra $\torus$. There is a canonical decomposition of the loop group
\begin{equation}
\label{declg}
LT\cong \Lambda\times V(S^1)\times T,
\end{equation}
where $V(S)=L\torus\slash\torus$, and the commutator pairing is given on the first and the third factor 
by the map $\Lambda\to \hat{T}$ defined by $\left<~,~\right>$.
The pairing on the middle factor is defined by means of the symplectic form induced by $\omega:L\torus\times L\torus\rightarrow\R$
\begin{equation}
\label{cocycle}
\omega(\xi,\eta):=\int_{S^1}\left<\xi,d\eta\right>,\quad \xi,\eta\in L\torus.
\end{equation}
This commutator pairing defines a a generalized Heisenberg group as defined in \S \ref{def-hg} with center given by $A\times\T$, where $A$ is the finite group 
$A:=\Lambda^\circ/\Lambda$ and 
\[
\label{dual-lattice}
\Lambda^\circ=\{\mu\in\torus,~\left<\mu,\lambda\right>\in\Z,~\forall\lambda\in\Lambda\}\cong\Hom(T,\T)
\]
is the dual lattice. To write down an explicit cocycle and thereby an explicit central extension,
we choose an integral bilinear form $B$ on $\Lambda$ such that 
\begin{equation}
\label{Z-2}
B(\lambda_1,\lambda_2)+B(\lambda_2,\lambda_1)=\left<\lambda_1,\lambda_2\right>,
\end{equation}
for all $\lambda_1,\lambda_2\in\Lambda$. This is possible because the lattice is even. With this,
the central extension of $LT$ is canonically a product of the Heisenberg extension of the symplectic 
vector space $V(S^1)$ and the extension of $T\times\Lambda$ given by the cocycle
\[
\psi\left((t_1,\lambda_1),(t_2,\lambda_2)\right):=(-1)^{B(\lambda_1,\lambda_2)}t_2^{\lambda_1}.
\]
There is a canonical polarization $V_\C(S^1)=V_+(S^1)\oplus V_-(S^1)$ by the decomposition into positive and negative Fourier modes, or, equivalently, the 
Hilbert transform. This defines a polarization class on $LT$ in the sense of Definition \ref{pol-ghg} which is invariant under the action of ${\rm 
Diff}^+(S^1)$. As shown in \cite[\S 9.5]{ps}, it is exactly this class that leads to positive energy representations in the sense of Definition \ref{per}.
By Theorem \ref{irgh}, the irreducible representations at level $q$ now correspond to the characters $\varphi\in\hat{A}$. To give an explicit construction, consider 
the unique Heisenberg representation $\h_{V(S^1)}$ of the Heisenberg extension of $V(S^1)$ defined by the polarization. Choose $\lambda\in\Lambda^\circ$, 
and let $T$ act on $\h_{V(S^1)}$ via $\lambda$. This yields an irreducible representation $\h_{\lambda}$ of the unit component $\widetilde{LT}_0$, and with this we define
\begin{displaymath}
\begin{split}
\h_{\varphi}:&=\Ind_{\widetilde{LT}_0}^{\widetilde{LT}}\left(\h_{\lambda}\right)\\
&=\bigoplus_{\mu\in\Lambda}\h_{\lambda+\mu}.
\end{split}
\end{displaymath} 
This clearly defines an irreducible representation of $\widetilde{LT}$ which only depends on 
$[\lambda]$, the image of $\lambda$ in $\hat{A}$. (From now we shall omit the brackets and just write $\lambda\in \hat{A}$.) Notice that only when the level turns $\Lambda$ into 
a unimodular lattice, the central extension of $LT$ is a Heisenberg group, i.e., has a unique 
irreducible representation.
\begin{remark}
\label{level-spin}
In the above we have assumed that the level, i.e., the quadratic form $q$ on $\pi_1(T)$, turns 
$\Lambda$ into an \textit{even} lattice. This condition has a topological origin: it is actually better 
to view the quadratic form  as a class $q\in H^1(T;\Z)\otimes H^1(T;\Z)$. Since $H^1(T;\Z)=H^2(BT;
\Z)$, and $BLT\simeq T\times BT$, we have $q\in H^3(BLT;\Z)$. It is this cohomology class 
that classifies, also for nonabelian compact Lie groups,  the central extensions that arise from 
positive energy representations.

Because of the homotopy equivalence $BLT\simeq LBT$, there is a transgression map 
\[
H^4(BT;\Z)\rightarrow H^3(BLT;\Z).
\]
The cohomology $H^*(BT;\Z)$ is generated by the first Chern class $c_1\in H^2(BT;\Z)$, 
and $H^4(BT;\Z)$ equals the space of integral even bilinear forms on $\pi_1(T)$. 
We therefore see that our condition on the quadratic form $q$ means that the corresponding class
in $H^3(BLT;\Z)$ is transgressed from $H^4(BT;\Z)$. It is this class in $H^4(BT,\Z)$ that is called the \textit{level} of the theory. 
For example, for $T=\T$, both cohomology groups are isomorphic to $\Z$, but the map turns out to be multiplication by $2$. 

The Conformal Field Theory (CFT) that we are about to describe comes from a Topological Quantum 
Field Theory (TQFT) in dimension 3, called abelian Chern--Simons theory, classified by $H^4(BT;\Z)$. From the point of view of loop groups, the structure defined by this theory, e.g., the modular 
tensor structure on the representation category, is therefore only defined for even levels, i.e., 
corresponding to inner products for which the lattice $\Lambda$ is even. To treat the odd level, 
corresponding to half-integer Chern--Simons theory, one needs to refine to a \textit{spin}-TQFT, in 
which manifolds are assumed to be equipped with a spin-structure. This TQFT is described in 
\cite{moore}. In this paper we will mainly deal with the case of an even level, but will remark where 
the spin structure comes in when treating the odd-level case. With this, the construction of the 
corresponding spin-CFT presents no difficulty.
\end{remark}
\subsection{The moduli space of holomorphic $T_\C$-bundles}
In this section we will describe how the theory of affine symplectic manifolds can be used to quantize certain moduli spaces $T$-bundles over surfaces with boundaries. Before turning to quantization, let us first describe this moduli space in some more detail, cf. \cite{ab,donaldson,mw}. There are essentially two approaches: the symplecto-geometric approach as the moduli space of flat $T$-bundles, and the complex analytic approach as the moduli space of holomorphic $T_\C$-bundles. The equivalence between the two provides the full K\"ahler structure on these moduli spaces.

Let $\Sigma$ be a compact Riemann surface, possibly with smooth boundaries.  Define
the moduli space $\M_T(\Sigma)$ as follows:
\[
\M_T(\Sigma):=\left\{\begin{array}{c}\mbox{Isomorphism classes of holomorphic $T_\C$-bundles $E$}\\ \mbox{with a smooth
trivialization $E|_{\partial\Sigma}\cong\partial\Sigma\times T_\C$.}
\end{array}\right\}
\]
This space turns out to have a natural complex manifold structure, that can be described as follows:
when $\partial\Sigma\neq\emptyset$, and $\Sigma$ contains no closed components, the interior is a Stein manifold, and consequently any holomorphic bundle is trivial. Trivializing all bundles over the interior therefore defines an isomorphism 
\begin{equation}
\label{donfactorization}
\M_T(\Sigma)\cong T_\C(\partial\Sigma)/T^\Sigma_\C,
\end{equation}
where $T^\Sigma_\C\subseteq T_\C(\partial\Sigma)$ is the closed subgroup of loops that admit a holomorphic extension to the surface $\Sigma$.
The K\"ahler structure of $\M_T(\Sigma)$ is slightly more involved, and follows from the following:
\begin{proposition}
\label{mfas}
The moduli space
$\M_T(\Sigma)$ is an affine symplectic manifold whose structure is independent of the 
complex structure on $\Sigma$.
\end{proposition}
\begin{proof}
This follows from the symplecto-geometric description of $\M_T(\Sigma)$. First remark that, although
$\M_T(\Sigma)$ is not connected, it has an abelian group structure by taking tensor products of bundles, and it therefore suffices to prove the statement of the unit component $\M^0_T(\Sigma)$ of $\M_T(\Sigma)$. 

Consider $\A(\Sigma)$, the space of connections on the trivial $T$-bundle over $\Sigma$. In two dimensions, this space carries a symplectic form 
\begin{equation}
\label{absymplectic}
\omega(\alpha,\beta)=\int_\Sigma\left<\alpha,\beta\right>,
\end{equation}
for $\alpha,\beta\in T_A\A(\Sigma)=\Omega^1(\Sigma,\torus)$, so that the pair $(\A(\Sigma),\omega)$ is an affine symplectic manifold. The action of the gauge group  $T(\Sigma)$ on $\calA(\Sigma)$, cf.\ equation \eqref{gauge}, preserves this symplectic form and is Hamiltonian with associated moment map given by $$\left<\mu(A),\xi\right>=\int_\Sigma\left<F_A,\xi\right>-\int_{\partial\Sigma}i_{\partial\Sigma}^*\left<A,\xi\right>,$$ for $\xi\in\Omega^0(\Sigma,\torus),$ where $i_{\partial\Sigma}:\partial\Sigma\hookrightarrow\Sigma$ is the canonical inclusion and $F_A=dA$ denotes the curvature of a connection. Let $T_\partial(\Sigma)$ be the subgroup of $T(\Sigma)$ consisting of gauge transformations which are trivial at the boundary. This gauge group fits into an exact sequence of the form 
\begin{equation}
\label{esgg}
1\rightarrow T_\partial(\Sigma)\rightarrow T(\Sigma)\rightarrow T(\partial\Sigma)\rightarrow H^2(\Sigma,\partial\Sigma;\Lambda)\rightarrow 0.
\end{equation}
Any connection defines a holomorphic structure on the trivial $T_\C$-bundle by the associated Cauchy--Riemann operator $\bar{\partial}_A$, and it was proved in \cite{donaldson} that this defines 
an isomorphism
\begin{equation}
\label{defmf}
\M^0_T(\Sigma)\cong\calA_F(\Sigma)\slash T_\partial(\Sigma),
\end{equation}
where $\calA_F(\Sigma)\subset\calA(\Sigma)$ is the subset of flat connections. 
The right hand side is a symplectic quotient, and this defines the weakly symplectic structure
on $\M_T(\Sigma)$. Let 
\begin{equation}
\label{vss}
V(\Sigma):=\left.\left\{\alpha\in\Omega^1(\Sigma,\torus),~d\alpha=0\right\}\right/d\left\{\beta\in\Omega^0(\Sigma,\torus),~\beta|_{\partial\Sigma}=0\right\}.
\end{equation}
 This vector space carries a symplectic form given by the formula \eqref{absymplectic}, and $V(\Sigma)$ acts on $\M^0_T(\Sigma)$, induced from 
the affine action $A\mapsto A+\alpha$ of $\Omega^1(\Sigma,\torus)$ on $\A(\Sigma)$. Notice that the gauge group $T_\partial(\Sigma)$ acts on 
$\calA(\Sigma)$ via the embedding $T_\partial(\Sigma)\hookrightarrow \Omega^1(\Sigma,\torus)$ given by $\varphi\mapsto -d\varphi\varphi^{-1}$. 
Since the Lie algebra of $T_\partial(\Sigma)$ is exactly given by $$\{\xi\in\Omega^0(\Sigma,\torus),~\xi|_{\partial\Sigma}=0\},$$ it follows from the 
definition \eqref{defmf} of $\M^0_T(\Sigma)$ that the isotropy groups $V(\Sigma)_{[A]}$, for $[A]\in\M^0_T(\Sigma)$, of the action of $V(\Sigma)$ 
on $\M^0_T(\Sigma)$ are given by $$V(\Sigma)_{[A]}=\pi_0(T_\partial(\Sigma))=H^1(\Sigma,\partial\Sigma;\Lambda).$$ This is a finitely 
generated lattice, proving that the moduli space $\M_T(\Sigma)$ is affine symplectic in the sense of Definition \ref{as}.
\end{proof}
\begin{remark}
\label{jacobian}
This description of $\M_T(\Sigma)$ remains true if $\partial\Sigma=\emptyset$. In this case, the moduli space $\M_T(\Sigma)$ is a disjoint union of finite dimensional symplectic tori \cite{ab}; 
$$
\M_T(\Sigma)\cong \Hom(\pi_1(\Sigma),T)\times H^2(\Sigma;\Lambda).
$$ 
For $T=\T$, the one-dimensional unitary group, this is nothing but the Jacobian of $\Sigma$ consisting of isomorphism classes of line bundles.
\end{remark}
\begin{corollary}
\label{fgms}
The homotopy groups of $\M_T(\Sigma)$ are given by $$\pi_0\left(\M_T(\Sigma)\right)\cong H^2(\Sigma,\partial\Sigma;\Lambda),\hspace{0.5cm}\pi_1\left(\M_T(\Sigma)\right)\cong H^1(\Sigma,\partial\Sigma;\Lambda),$$ and $\pi_n(\M_T(\Sigma))=0$ for $n\geq 2$.
\end{corollary}
The above description of $\M_T(\Sigma)$ as a symplectic quotient shows that the natural 
action of the boundary gauge group $T(\partial\Sigma)$ by changing the boundary framing,
preserves the symplectic form and is Hamiltonian
with moment map $\mu_\Sigma$ given by 
\begin{equation}
\label{mm}
[A]\mapsto -A|_{\partial\Sigma},
\end{equation}
for $[A]\in\M_T(\Sigma)$. Acting on the unit element in $\M_T(\Sigma)$, i.e., the trivial $T_\C$-bundle,
we have:
\begin{lemma}
\label{esms}
The moduli space and the gauge group fit into an exact sequence of groups
\[
1\rightarrow H^0(\Sigma,T)\rightarrow T(\partial\Sigma)\rightarrow\M_T(\Sigma)\rightarrow H^1(\Sigma,T)\rightarrow 1.
\]
\end{lemma}
\begin{proof}
The proof of this lemma is easy once one realizes that $H^1(\Sigma,T)$ is the moduli space of flat $T$-bundles on $\Sigma$ and the map $\M_T(\Sigma)\rightarrow H^1(\Sigma,T)$ forgets the boundary framing.
\end{proof}
\begin{proposition}
\label{polcl}
The complex structure on $\Sigma$ defines a polarization of $\M_T(\Sigma)$. In fact, any other 
complex structure defines a polarization in the same class.
\end{proposition}
\begin{proof}
A complex structure on $\Sigma$ turns $\A(\Sigma)$ into a complex K\"{a}hler manifold by the Hodge $*$-operator $*:\Omega^1(\Sigma,\torus)\rightarrow\Omega^1(\Sigma,\torus)$ satisfying $*^2=-1$, with associated K\"ahler metric given by $$Q(\alpha,\beta)=\int_\Sigma\left<\alpha,*\beta\right>,$$ which clearly shows that the complex structure is compatible with the symplectic form. Since $\M_T(\Sigma)$ is the symplectic quotient of $\A(\Sigma)$ with respect to the action of $T_\partial(\Sigma)$, cf.\  equation \eqref{defmf}, this induces a compatible complex structure on $\M_T(\Sigma)$ by Proposition \ref{sr}, which is the same as that induced by the isomorphism \eqref{donfactorization}. 

On the other hand, $\M_T(\Sigma)$ carries a canonical polarization class induced from the one on $T(\partial\Sigma)$ by Lemma \ref{esms}. As in \cite[\S 8.11]{ps}, it follows that the polarization induced by a complex structure lies exactly in this same polarization class. This completes the proof.
\end{proof}

\subsubsection{The prequantum line bundle}
So far, we have constructed the moduli spaces $\M_T(\Sigma)$ as a polarized affine symplectic manifold. However, we want to consider them additionally as being equipped with a Hamiltonian $T(\partial\Sigma)$-action. This contains slightly more data, since, contrary to the components $V(\partial\Sigma)\times H^0(\partial\Sigma,\Lambda)$ in the decomposition \eqref{declg}, the  subgroup $H^0(\partial\Sigma,T)\subset T(\partial\Sigma)$ does not quite act in an affine way. 
Any prequantum line bundle $L$ defines a central extension of the loop group $T(\partial\Sigma)$. Indeed, since $T(\partial\Sigma)$ acts in a Hamiltonian fashion, its action preserves the isomorphism class of $L$, and the covering automorphisms 
define a central extension of $T(\partial\Sigma)$. The Lie algebra cocycle of this central extension is given by the symplectic form, and it follows from Stokes' theorem that 
\begin{displaymath}
\begin{split}
\omega(v_\xi,v_\eta))=\int_{\Sigma}\left<d\xi,d\eta\right>=\int_{\partial\Sigma}\left<\xi,d\eta\right>
\end{split}
\end{displaymath}
for $\xi,\eta\in \Omega^0(\Sigma,\torus)={\rm Lie}(T(\Sigma))$ with generating vector fields $v_\xi,v_\eta$. Since the right hand side is exactly the fundamental cocycle \eqref{cocycle}, the induced central extension must be isomorphic to the central extension of \S \ref{iper} associated to the inner product $\left<~,~\right>$ on $\torus$. Next, consider the set ${\rm Preq}(\Sigma)$ of isomorphism classes of 
of prequantum line bundles on $\M_T(\Sigma)$, considered as equivariant line bundles.
\begin{proposition}
\label{picard-pq}
There is a short exact sequence 
$$
0\rightarrow H_0(\Sigma;\hat{A})\rightarrow {\rm Preq}(\Sigma)\rightarrow H_1(\Sigma,\partial\Sigma;\hat{\Lambda}) \rightarrow 0.
$$
\end{proposition}
\begin{proof}
The third arrows is given by the forgetful map where one forgets about the $T(\partial\Sigma)$-action and considers $\M_T(\Sigma)$ solely as an affine symplectic manifold: remark that the Pontryagin dual of $\pi_1(\M_T(\Sigma))=H^1(\Sigma,\partial\Sigma;\Lambda)$ is given by $H_1(\Sigma,\partial\Sigma;\hat{\Lambda})$. 

The second map is given by the following construction, following \cite{mw}, of a prequantum line bundle $L(\Sigma)\rightarrow\M_T(\Sigma)$.  On $\calA(\Sigma)$, consider the trivial line bundle $L:=\calA(\Sigma)\times\C$ 
with its canonical metric and connection 
$$
\nabla_\alpha(s)(A)=ds(A)-\sqrt{-1}
\int_\Sigma\left<\alpha,A\right>,
$$ 
for $\alpha\in T_A\calA(\Sigma)=\Omega^1(\Sigma,\torus)$, 
and $s:\calA(\Sigma)\rightarrow\C$ a smooth section. One easily finds that $F_\nabla=-\sqrt{-1}\omega$, 
i.e., $(L,\nabla)$ defines a prequantum line bundle on $\calA(\Sigma)$. The group cocycle $c:T(\Sigma)\times T(\Sigma)\rightarrow \T$ defined by 
\begin{equation}
\label{ggrc}
c(\varphi_1,\varphi_2)=\exp\left(-\sqrt{-1}\pi\int_\Sigma\left<\varphi_1^{-1}d\varphi_1,d\varphi_2\varphi^{-1}_2\right>\right)
\end{equation}
defines a central extension $\widetilde{T(\Sigma)}$ of $T(\Sigma)$ which naturally acts on $L$ by $$(\varphi,z)\cdot (A,w)=(\varphi\cdot A,\exp\left(\sqrt{-1}\pi\int_\Sigma\left<\varphi^{-1}d\varphi,A\right>\right)zw).$$
Next, observe that the central extension of $T(\Sigma)$ can be trivialized over the subgroup 
$T_\partial(\Sigma)$: by Stokes' theorem the cocycle \eqref{ggrc} is canonically trivial over the 
identity component of $T_\partial(\Sigma)$. The induced central extension of the group of 
components $\pi_0(T_\partial(\Sigma))=H^1(\Sigma,\partial\Sigma;\Lambda)$ is defined by the 
cocycle $\exp(\sqrt{-1}\pi Z)$ 
where $Z$ is the antisymmetric bilinear form $Z(\psi_1,\psi_2)=\left<\psi\cup\psi_2,[\Sigma]\right>,$ for $\psi_1,
\psi_2\in H^1(\Sigma,\partial\Sigma;\Lambda)$ and the cup-product also includes the inner product 
on $\Lambda$. Since the lattice is assumed to be even, this cocycle is trivial.  With this trivialization,
we define the line bundle $L^0(\Sigma)$ over the unit component $\M_T^0(\Sigma)$ as
\[
L^0(\Sigma)=L|_{\calA_F(\Sigma)}\slash T_\partial(\Sigma).
\]
Clearly, this line bundle carries an action of the extension 
$$
1\rightarrow H^0(\Sigma,\T)\rightarrow\widetilde{T(\Sigma)}/T_\partial(\Sigma)\rightarrow T(\partial\Sigma).
$$ 
In fact, this is a central extension of the kernel of the map $T(\partial\Sigma)\rightarrow H^2(\Sigma,\partial\Sigma;\Lambda)$ given by \eqref{esgg}. It remains 
to extend the line bundle to the other connected components of $\M_T(\Sigma)$. 
As an abelian group, all components are isomorphic to $\M_T^0(\Sigma)$, so the 
line bundle $L(\Sigma)$ is uniquely defined by requiring the quotient group 
$T(\partial\Sigma)\slash T(\Sigma)\cong H_0(\Sigma;\Lambda)=\pi_0(\M_T(\Sigma))$ to act by isomorphisms.  Finally, since $H^0(\Sigma,T)$ acts trivially on $\M_T(\Sigma)$ by the restriction 
$H^0(\Sigma,T)\to H^0(\partial\Sigma,T)$, an element $\lambda\in H^0(\Sigma,\Lambda^\circ)$
defines a character of $H^0(\Sigma,T)$ and thereby a lift of the trivial action on $M_T(\Sigma)$ to $L(\Sigma)$: it acts by the character $\lambda+\chi$ over the connected component labeled by 
$\chi\in H_0(\Sigma;\Lambda)$. This construction defines a map $H_0(\Sigma;\Lambda^\circ)\to
{\rm Preq}(\Sigma)$, which clearly factors over the quotient by $H_0(\Sigma;\Lambda)$. This defines the second map. Clearly, the sequence is exact.
\end{proof}
From now on, we shall restrict our attention to the prequantum line bundle obtained from the 
unit element $0\in H_0(\Sigma;\hat{A})$, and write $L(\Sigma)$ for this line bundle. This is the line bundle on which $H^0(\Sigma;T)$ acts by the trivial representation on the unit component of 
$\M_T(\Sigma)$. There are other constructions of this line bundle than the one we have given above: one is to use the $\C^*$-bundle of the central extension of $T_\C(\partial\Sigma)$ together with fact that by Cauchy's theorem, this central extension is canonically trivial over $T^\Sigma_\C$, and the isomorphism \eqref{donfactorization}. Another construction is to use the determinant line bundle
of the Cauchy--Riemann operator $\bar{\partial}_A$ associated to a connection $A\in\calA(\Sigma)$.
This last description shows how the line bundle extends when the complex structure is allowed to vary: the fiber of the unit element in $\M_T(\Sigma)$ forms a line bundle that is isomorphic to $\operatorname{det}_\Sigma^{\otimes c}$, where $c=\dim(T)$, and $\operatorname{det}_\Sigma$  
denotes the determinant line bundle of the ordinary $\bar{\partial}$-operator.
\begin{remark}
In the proof above we used the fact that the lattice $\Lambda$ is assumed to be even to define a splitting of a central extension over $T_\partial(\Sigma)$. When the level is odd, one can only define
a splitting after choosing a cobounding spin structure on $\Sigma$ and in this case the exact sequence 
is given by  
\[
0\rightarrow H_0(\Sigma;\hat{A})\times H_1(\Sigma,\partial\Sigma;\Z/2\Z)\rightarrow {\rm Preq}(\Sigma)\rightarrow H_1(\Sigma,\partial\Sigma;\hat{\Lambda}) \rightarrow 0,
\]
where the factor $H_1(\Sigma,\partial\Sigma;\Z/2\Z)$ labels isomorphism classes of spin structures 
on $\Sigma$. This is clear from the complex point of view in equation \eqref{donfactorization} as 
follows: as a $\C^*$-bundle, a prequantum line bundle is induced by the complex central extension
of $T_\C(\partial\Sigma)$, as proved in \cite[Prop. 12.4]{segal} a spin structure on $\Sigma$ defines 
a splitting of the central extension restricted to $T^\Sigma_\C$ for odd levels.
\end{remark}

\subsection{Quantization} Now that we have defined a line bundle $L(\Sigma)$ over $\M_T(\Sigma)$, we pick a complex structure on $\Sigma$ and define $\h_{\Sigma}$ to be the quantization as defined in Section \ref{squantization} associated to these data. Explicitly, one has 
\[
\mathcal{H}_{\Sigma}=\bigoplus_{\chi\in H_0(\Sigma,\Lambda)}\h_{\M^\chi_T(\Sigma)},
\]
 where the direct sum is taken in the $L^2$-sense. When $\partial\Sigma=\emptyset$, 
we denote by $\h_\Sigma$ the quantization of the unit component, which corresponds to the $H^0(\Sigma,T)$-invariants of a Hilbert space sum
as above. (This is due to the fact that the moduli \textit{stack} of $T_\C$-bundles contains a factor $BT_\C$ when $\Sigma$ is closed, a fact that will not be pursued further in this paper.)
From \S \ref{quantization} we know that each  $\h_{\M^\chi_T(\Sigma)}$ carries an irreducible representation of a generalized Heisenberg group associated to $\M^\chi_T(\Sigma)$. Let 
us analyze this group in some more detail. First notice that the linearization of Lemma \ref{esms} yields an exact sequence 
\[
0\rightarrow H^0(\Sigma,
\torus)\rightarrow\Omega^0(\partial\Sigma,\torus)\rightarrow V(\Sigma)\rightarrow H^1(\Sigma,\torus)\rightarrow 0,
\] 
where the fourth map is induced by taking 
the de Rham cohomology class of a closed differential form, cf.\ definition \eqref{vss}. By definition, $V_{\M_T^0(\Sigma)}=H^1(\Sigma,\partial\Sigma;\Lambda)^\perp$ in the Heisenberg group associated to $V(\Sigma)$, and therefore we have the exact sequence
\begin{equation}
\label{ex-hg}
1\to H^0(\Sigma,T)\to T(\partial\Sigma)_0\to V_{\M_T^0(\Sigma)}\to H^1(\Sigma;\Lambda^\circ)\to 1.
\end{equation}
In this exact sequence, the third map pulls back the cocycle on $V_{\M^0_T(\Sigma)}$ to the cocycle
on $T(\partial\Sigma)_0=V(\partial\Sigma)\times H^0(\partial\Sigma,T)$ defining the generalized Heisenberg extensions. This describes the structure of each of the summands $\h_{\M_T^\chi(\Sigma)}$. We can extend the representation of the Heisenberg group of $V_{\M_T^0(\Sigma)}$ on each of these summands to an irreducible representation of a generalized Heisenberg group on the whole $\h_\Sigma$ by defining
\[
\mathcal{G}(\Sigma):=V_{\M_T^0(\Sigma)}\times \left(H^0(\Sigma,T)\times H_0(\Sigma,\Lambda)\right).
\]
As before, the inner product on $\torus$ defines a map $\Lambda\to\hat{T}$ and thereby a cocycle on the second component of this abelian group. Together with the restriction of the symplectic form \eqref{absymplectic} on the fist component,
this defines a generalized Heisenberg group $\widetilde{\mathcal{G}(\Sigma)}$ that acts irreducibly on $\h_\Sigma$. Clearly, we have
\[
Z\left(\mathcal{G}(\Sigma)\right)=H^1(\Sigma,\partial\Sigma;\Lambda)\times H^0(\Sigma;A),
\] 
which naturally fits with the computation of the Picard group in Proposition \ref{picard-pq}. 
If we take the quotient by this center to define $\mathcal{G}(\Sigma)_{red}:=\mathcal{G}(\Sigma)\slash Z\left(\mathcal{G}(\Sigma)\right)$, it follows from \eqref{ex-hg} that this group fits into an exact sequence
\begin{equation}
\label{fund-seq}
1\to H^0(\Sigma;A)\to T(\partial\Sigma)\to \mathcal{G}(\Sigma)_{red}\to H^1(\Sigma;A)\to 1.
\end{equation}
The crucial point, and the connection to the representation theory of loop groups, is the following:
\begin{proposition}
The Hilbert space $\h_{\Sigma}$ carries a positive energy representation of $T(\partial\Sigma)$.
\end{proposition}
\begin{proof}
As we have seen, the line bundle $L(\Sigma)\rightarrow\M_T(\Sigma)$ carries a natural action of the  
central extension of $T(\partial\Sigma)$ defined by the inner product $\left<~,~\right>$ by 
automorphisms. This central extension therefore acts projectively on the associated space of 
holomorphic sections. The affine part of $T(\partial\Sigma)$, i.e., $V(\partial\Sigma)$, acts via the 
morphism in the exact sequence \eqref{ex-hg}, and will therefore be implemented by a projective 
unitary representation on $\h_\Sigma$. The extension to $T(\partial\Sigma)$ is projective unitary as 
well, because the geometric action of the Heisenberg extension of $H^0(\partial\Sigma,T)\times H_0
(\partial\Sigma,\Lambda)$ preserves the $L(\Sigma)$-valued measure determined by the complex 
structure.

Finally let us prove that this representation has positive energy. As observed in the proof of 
Proposition \ref{mfas}, the standard polarization $V(\partial\Sigma)=V_+(\partial\Sigma)\oplus 
V_-(\partial\Sigma)$ of $V(\partial\Sigma)$ induces a polarization class of $\M_T(\Sigma)$ by 
Lemma \ref{esms}, which equals the class defined by any complex structure on $\Sigma$.  But, as in 
\S 2.1, this standard polarization yields a positive energy representation, and therefore, it follows 
from Theorem \ref{irgh} that $\h_{\Sigma}$ is of positive energy.
\end{proof}
Let $\e(\Sigma)$ be the moduli space of Riemann surfaces with parameterized boundaries that are topologically equivalent to $\Sigma$. Each point in $\e(\Sigma)$ defines a polarization of $\M_T(\Sigma)$ and thereby a quantization $\h_\Sigma$: this is the unique irreducible representation $\h_\Sigma$ of the generalized Heisenberg group of $\G(\Sigma)$ equipped with a holomorphic, $\C^*$-linear isometric map $L(\Sigma)\to\h_\Sigma$. The associated bundle of projective spaces is canonically trivial over $\e(\Sigma)$ and restricted to the trivial bundle in $\M_T(\Sigma)$, one finds a map $\e(\Sigma)\to\proj(\h_\Sigma)$.
\begin{theorem}
\label{mpcg}
The projective unitary representation of $T(\partial\Sigma)$ on $\h_{\Sigma}$ extends to the semi-direct product $\mathcal{K}_\Sigma\ltimes T(\partial\Sigma)$, where $\mathcal{K}_\Sigma$ is an extension of {\rm Diff}$^+(\partial\Sigma)$ by the mapping class group $\Gamma(\Sigma,\partial\Sigma)$; $$1\rightarrow\Gamma(\Sigma,\partial\Sigma)\rightarrow\mathcal{K}_\Sigma\rightarrow {\rm Diff}^+(\partial\Sigma)\rightarrow 1.$$
 In particular, $\h_\Sigma$ carries a projective unitary representation of $\Gamma(\Sigma,\partial\Sigma)$ commuting with $LT$.
\end{theorem}
\begin{proof}
The natural action of the group Diff$^+(\Sigma)$ of orientation-preserving diffeomorphisms of $\Sigma$ on $\M_T(\Sigma)$ factors over Diff$_0^+(\Sigma,\partial\Sigma)$, the identity component of the subgroup of diffeomorphism which leave the boundary point-wise fixed, as one easily sees from Lemma \ref{esms}.  In view of the definition \eqref{defmf} of $\M_T(\Sigma)$, this induces an action of $$\mathcal{K}_\Sigma:={\rm Diff}^+(\Sigma)/{\rm Diff}_0^+(\Sigma,\partial\Sigma).$$  Clearly, the action of Diff$^+(\Sigma)$ preserves the symplectic form \eqref{absymplectic}, and one finds a natural map $\mathcal{K}_\Sigma\rightarrow{\rm Aut}_{res}(\M_T(\Sigma))$. The desired representation now follows from Prop. \ref{aut}. The exact sequence in statement of the proposition follows immediately from the definition $\Gamma(\Sigma,\partial\Sigma):={\rm Diff}^+(\Sigma,\partial\Sigma)/{\rm Diff}^+_0(\Sigma,\partial\Sigma)$ and the fact that ${\rm Diff}^+(\Sigma)/{\rm Diff}^+(\Sigma,\partial\Sigma)\cong {\rm Diff}^+(\partial\Sigma)$.
\end{proof}
\begin{example}  
\label{ex-quant}
Well known positive energy representations of $LT$ are special cases of the quantization procedure described above:
\begin{itemize}
\item[$i)$] Let $\Sigma=D$ be a disk. Then we have $\M_T(D)\cong LT/T$, and the quantization yields $\h_D=\h_0$, the basic representation at the given level. In this case, Proposition \ref{picard-pq} gives 
$
{\rm Preq}(D)=\hat{A},
$ and if we use the line bundle $L_\varphi$ labeled by $\varphi\in \hat{A}$, we obtain all other irreducible representations of $LT$ as in \S \ref{iper}. Theorem \ref{mpcg} now gives the well known
fact that these representation extend to the semi-direct product ${\rm Diff}(S^1)\ltimes LT$.
\item[$ii)$] For $\Sigma=A$ an annulus, we easily see from the exact sequence \eqref{fund-seq} that there is an isomorphism
\begin{equation}
\label{sf}
\h_A\cong\bigoplus_{\lambda\in \hat{A}}\h_{\lambda}\otimes\h^*_{\lambda}
\end{equation}
 of projective $LT\times LT^{\mbox{\tiny op}}$-representations. This isomorphism is even canonical: 
 the moduli space $\e(A)$ forms a semigroup under gluing of annuli and the line given by the fiber of the prequantum line bundle over the trivial bundle in $\M_T(A)$ defines a central extension of it. It is well-known, cf.\ \cite{segal} that this central extension acts on each $\h_\lambda$ by trace class operators. 
 This defines a vector $\Omega_A$ in the Hilbert space on the right hand side and, by Theorem \ref{qbundle} fixes a canonical isomorphism. In  the case of $A_q:=\{z\in\C,~q\leq |z|\leq 1\}$ for $q\in(0,1)$, the operator is simply given by $q^D$, where 
  $D$ is the generator of the $\T$-representation that defines the energy grading on $\h_\lambda$.\end{itemize}
\end{example}

\subsection{Gluing and reduction} 
Let $\Sigma_1$ and $\Sigma_2$ be two Riemann surfaces with parametrized boundaries. We can use the parametrization to glue $\Sigma_1$ and $\Sigma_2$ over a compact $1$-manifold $S$, of course diffeomorphic to a disjoint union of copies of $S^1$. We denote the resulting smooth surface by $\Sigma=\Sigma_1\cup_S\Sigma_2$. Its boundary, which may be empty, inherits a canonical parametrization from $\Sigma_1$ and $\Sigma_2$. 
\begin{proposition}
\label{ger}
{\rm (cf.\ \cite[Thm. 3.5]{mw})} In the situation above, the moduli space $\M_T(\Sigma)$ is obtained from the moduli spaces associated to $\Sigma_1$ and $\Sigma_2$ by symplectic reduction; $$\M_T(\Sigma)\cong\Sr{\left(\M_T(\Sigma_1)\times\M_T(\Sigma_2)\right)} T(S).$$ The same holds true for the prequantum line bundle; $$ L(\Sigma)\cong\Sr{\left(L(\Sigma_1)\times L(\Sigma_2)\right)} T(S).$$
\end{proposition}
\begin{proof}
Consider first the connected component of the unit $T(S)_0$ acting on $\M_T^0(\Sigma_1)\times\M^0_T(\Sigma_2)$. Since the moment map of the action of $T(S)$ on $\M^0_T(\Sigma_1)$ and $\M^0_T(\Sigma_2)$ is given by restriction of connections to the boundary, cf.\ \eqref{mm}, the zero locus of the moment map on $\M^0_T(\Sigma_1)\times\M^0_T(\Sigma_2)$ corresponding to the diagonal $T(S)_0$-action is given by the set $([A_1],[A_2])\in \M^0_T(\Sigma_1)\times\M^0_T(\Sigma_2)$ satisfying $$A_1|_{\partial\Sigma_1}=A_2|_{\partial\Sigma_2}.$$ As in \cite[Thm 3.5]{mw}, one shows that this defines, up to the $T(S)_0$-action, a unique gauge equivalence class of a smooth flat connection $[A]$ on $\Sigma$. This defines a smooth symplectomorphism 
\[
\Sr{\left(\M^{0}_T(\Sigma_1)\times\M^{0}_T(\Sigma_2)\right)}{T(S)_0}\stackrel{\cong}{\longrightarrow}\M_T^0(\Sigma).
\]
Extending the action to the other connected component gives isomorphisms
\[
\Sr{\left(\M^{\chi_1}_T(\Sigma_1)\times\M^{\chi_2}_T(\Sigma_2)\right)}{T(S)_0}\stackrel{\cong}{\longrightarrow}\M_T^{\chi_1-\chi_2}(\Sigma),
\]
where the map $(\chi_1,\chi_2)\mapsto\chi_1-\chi_2\in H_0(\Sigma;\Lambda)$ is the natural morphism
appearing in the Mayer--Vietoris sequence, cf.\ Remark \ref{mv} below. By exactness of this sequence,
this morphism is surjective with kernel given by $H_0(S;\Lambda)=\pi_0(T(S))$. Acting by the full loop group $T(S)$ instead of the unit component therefore precisely gives the isomorphism of the theorem.
This proves the statement for the moduli spaces. The gluing of the prequantum line bundle is proved in similar fashion; we omit the details.
\end{proof}
\begin{remark}
\label{mv}
As affine symplectic manifolds, the modelling spaces of the manifolds in this proposition are related by symplectic reduction of vector spaces:
\begin{equation}
\label{lsrms}
V(\Sigma)\cong\left(V(\Sigma_1)\times V(\Sigma_2)\right)\sr V(S).
\end{equation}
Using Corollary \ref{fgms} and applying Poincar\'e--Lefschetz duality, the exact sequence \eqref{esfg} of fundamental groups amounts to the Mayer--Vietoris sequence in homology:
\begin{equation}
\label{mvgl}
\begin{split}
0 \rightarrow H_1(S;\Lambda)&\rightarrow H_1(\Sigma_1;\Lambda)\oplus H_1(\Sigma_2;\Lambda)\rightarrow H_1(\Sigma;\Lambda)\rightarrow\\& \rightarrow H_0(S;\Lambda)\rightarrow H_0(\Sigma_1;\Lambda)\oplus H_0(\Sigma_2;\Lambda)\rightarrow H_0(\Sigma;\Lambda)\rightarrow 0.
\end{split}
\end{equation}
(In comparison to \eqref{esfg}, there is an extra term corresponding to $\pi_1(T(S))=H^0(S,\Lambda)$ since the loop group $T(S)$ is not quite affine; it contains a factor $H^0(S,T)$.) 
\end{remark}
\begin{remark}
As explained in \cite{segal}, a complex structure on $\Sigma_1$ and $\Sigma_2$ determines one on $\Sigma$. In fact, this defines a holomorphic map 
\[
\e(\Sigma_1)\times\e(\Sigma_2)\to\e(\Sigma)
\]
which induces the polarization of $\M_T(\Sigma)$ as an affine symplectic manifold implied by 
Proposition \ref{sr}. Identifying the prequantum line bundle $L(\Sigma)$ as a determinant line bundle,
the isomorphism of line bundles reduces to the isomorphism
\begin{equation}
\label{det}
\Det_{\Sigma_1}\otimes\Det_{\Sigma_2}\cong\Det_\Sigma
\end{equation}
proved in \cite[\S 6]{segal}.
\end{remark}
We now turn to the quantum version of this theorem. Remarkably, this involves the following
version of the gluing of surfaces in which the one-manifold $S$ is ``thickened up'' to a complex 
annulus $A$ with parameterized boundaries. We shall write $\Sigma_1\cup_A\Sigma_2$ 
for the gluing of $\Sigma_1$ and $\Sigma_2$ along such an annulus.
\begin{theorem}
\label{qcr}
Let $\Sigma_1$ and $\Sigma_2$ be two Riemann surfaces with parameterized boundaries.
There exists a canonical isometry
\[
 \h_{\Sigma_1\cup_A\Sigma_2}\cong\bigoplus_{\lambda\in H^0(S,\hat{A})}\Hom_{\widetilde{T(S)}}\left(\h_{\lambda}, \h_{\Sigma_1}\right)
 \otimes \Hom_{\widetilde{T(S)}_{op}}\left(\h^*_\lambda,\h_{\Sigma_2}\right).
 \]
\end{theorem}
\begin{proof}
The strategy of the proof is straightforward: we first show that the right hand side carries an
irreducible representation of the Heisenberg extension $\G(\Sigma)$ with the right character,
and secondly that there exists a canonical isometric embedding of the prequantum line bundle
$L(\Sigma_1\cup_A\Sigma_2)$. Theorem \ref{qbundle} then gives the desired isomorphism.

Next, using \eqref{sf}, let us rewrite the right hand side as
\[
\Hom_{\widetilde{T(S)}\times\widetilde{T(S)}_{op}}\left(\h_A,\h_{\Sigma_1}\otimes\h_{\Sigma_2}\right).
\]
The commutant of $\widetilde{T(S)}\times\widetilde{T(S)}_{op}$ in $\widetilde{A(\Sigma_1)}\times\widetilde{A(\Sigma_2)}$ clearly acts unitarily on this Hilbert space by composition
with intertwiners. Let us now consider, for notational simplicity only, the case that $S=S^1$, and 
remark that we can decompose
\[
\h_A\cong\h_V\otimes\h_V^*\otimes L^2(T\times\Lambda),
\]
using the decomposition \eqref{declg} of $LT$,
where $L^2(T\times\Lambda)$ is the ``regular representation'' of the generalized Heisenberg extension
of $T\times\Lambda$. Indeed this Hilbert space decomposes as
\[
L^2(T\times\Lambda)\cong \bigoplus_{\lambda\in\hat{A}}V_\varphi\otimes V^*_\varphi,
\]
where $V_\varphi$ is the irreducible representation of $\widetilde{T\times\Lambda}$ 
in which the center $A$ acts via the character $\lambda\in\hat{A}$.
Before proceeding with the proof, consider the following:
\begin{lemma}
\label{ff}
Let $\h$ carry a representation of $\widetilde{(\Lambda\times T)}\times \widetilde{(\Lambda\times T)}_{op}$. There is a canonical isomorphism $$\Hom_{\widetilde{(\Lambda\times T)}\times\widetilde{(\Lambda\times T)}_{op}}\left(L^2\left(\Lambda\times T\right),\h\right)\stackrel{\cong}{\longrightarrow}\h^{T\times T}.$$ 
\end{lemma}
\begin{proof}
Let $f_0\in L^2(\Lambda\times T)$ be the function which is $1$ on $\{0\}\times T$ and zero else. This defines a map $$\Hom_{\widetilde{(\Lambda\times T)}\times\widetilde{(\Lambda\times T)}_{op}}\left(L^2\left(\Lambda\times T\right),\h_1\otimes\h_2\right)\rightarrow\h_1\otimes\h_2$$ by $\psi\mapsto \psi(f_0)$. This map is clearly an isometry because 
\begin{displaymath}
\begin{split}
||\psi(f_0)||^2&=\left<\psi(f_0),\psi(f_0)\right>\\
&=\left<\psi^*\psi(f_0),f_0\right>\\
&=\psi^*\psi\left<f_0,f_0\right>\\
&=||\psi||^2,
\end{split}
\end{displaymath}
where we have used Schur's lemma. Since $f_0$ is invariant under the diagonal $T\times T$-action, it maps into $\h^{T\times T}.$ It is easy to see that it is surjective onto this space, it therefore induces the unitary isomorphism of the lemma.
\end{proof}
Consider now first the intertwiner space
\[
\Hom_{\widetilde{V(S)}\times\widetilde{V(S)}_{op}}\left(\h_{V(S)}\otimes\h^*_{V(S)},\h_{\M_T^0(\Sigma_1)}\otimes\h_{\M^0_T(\Sigma_2)}\right).
\]
By Theorem \ref{qcras}, this Hilbert space is isomorphic to the quantization
of the ``partial reduction''
\[
\M^{part,0}_T(\Sigma):=\left(\M^0_T(\Sigma_1)\times\M^0_T(\Sigma_2)\right)\sr V(S),
\]
an affine symplectic manifold modelled on $V(\Sigma)$, by Proposition \ref{sr} and \eqref{lsrms}.
Indeed, observe that $\pi_1(\M^{part,0}_T(\Sigma))=H_1(\Sigma_1;\Lambda)\oplus H_1(\Sigma_2;\Lambda)$ and that the intertwiner space above carries a representation
of the commutant of $H_1(\Sigma_1;\Lambda)\oplus H_1(\Sigma_2;\Lambda)$ in the Heisenberg group of $V(\Sigma)$. To see that this representation is irreducible one uses induction, cf.\ Proposition \ref{indaf}, together with the fact that \eqref{lsrms} gives $\h_{V(\Sigma_1)}\otimes \h_{V(\Sigma_2)}\cong \h_{V(S)}\otimes\h^*_{V(S)}\otimes\h_{V(\Sigma)}$: this shows that the induced representation is irreducible.

In view of the Lemma above, it remains to take the invariant with respect to the action of $H^0(S\sqcup S,T)$. Let us first consider the ``diagonal'' action since this action comes from the geometric action on
$\M^0_T(\Sigma_1)\times\M^0_T(\Sigma_2)$. Since $H^0(S,T)$ is compact, the invariant subspace carries an irreducible representation of the commutant of $H_1(\Sigma_1;\Lambda)\oplus H_1(\Sigma_2;\Lambda)$ {\em and} $H^0(S,T)$ in the Heisenberg group of $V(\Sigma)$.
But this is exactly the commutant of $H_1(\Sigma;\Lambda)$ by the Mayer--Vietoris sequence \eqref{mvgl}. Finally, the anti-diagonal copy of $H^0(S,T)$ is dual to the group of components $H_0(S;\Lambda)$ and therefore taking the invariants amounts to modding out the lattice $H_0(S;\Lambda)$
in $H_0(\Sigma_1;\Lambda)\oplus H_0(\Sigma_2;\Lambda)$: again by \eqref{mvgl} this gives $H_0(\Sigma;\Lambda)$. This shows that the intertwiner space of the theorem carries an irreducible representation of the generalized Heisenberg group $\G(\Sigma)$ associated to $\M_T(\Sigma)$. 

Finally, consider the embedding of $L(\Sigma_1\cup_A\Sigma_2)$. First recall, cf.\ Proposition \ref{qbundle}, that there are embeddings $L(\Sigma_i)\backslash \{0\}\to\h_{\Sigma_i}$ described dually
by $l_{\Sigma_i}\mapsto ev_{l_{\Sigma_i}}\in\h^*_{\Sigma_i}$, $i=1,2$. By Proposition \ref{ger}, we have
\[
L(\Sigma_1\cup_A\Sigma_2)\cong \Sr{\left(L(\Sigma_1)\times L(A)\times L(\Sigma_2)\right)}{\left(T(S)\times T(S)\right)}.
\]
With this, represent an element in $[l]\in L(\Sigma_1\cup_A\Sigma_2)$ by a triple $l=(l_{\Sigma_1},l_A,l_{\Sigma_2})$, unique up to the $LT\times LT$-action.
Consider the following linear functional
\[
ev_l:\Hom_{\widetilde{T(S)}\times\widetilde{T(S)}_{op}}\left(\h_A,\h_{\Sigma_1}\otimes\h_{\Sigma_2}\right)\to\C,
\]
defined by
\begin{equation}
\label{ev}
ev_l(\psi):=\left<l_{\Sigma_1}\otimes l_{\Sigma_2},\psi(l_A)\right>,
\end{equation}
where $l_A\in L(A)\subset\h_A$.
Clearly, since $\psi$ is an intertwiner, this functional only depends on the equivalence class
$[l]\in L(\Sigma_1\cup_A\Sigma_2)$.
It is easy to check, using the fact that the canonical map \eqref{can_iso} is an isometry, that this
defines an isometric embedding of $L(\Sigma_1\cup_A\Sigma_2)$. This completes the proof.
\end{proof}
Next, consider a ``partial labeling'' of the boundary of a Riemann surface $\Sigma$ with parameterized boundaries: we write $\partial\Sigma=\partial\Sigma_{f}\sqcup\partial\Sigma_{l}$, where $\partial\Sigma_f$ is the ``free'' part of the boundary, and we choose $\vec{\varphi}\in H^0(\partial\Sigma_l;\hat{A})$. Associated with this choice is the affine symplectic manifold
\[
\M_T(\Sigma\cup_{\partial\Sigma_l}D)\cong\Sr{\left(\M_T(\Sigma)\times\M_T(D)\right)}{T(\partial\Sigma_l)},
\]
where $D$ is a union of unit disks in the complex plane that are glued along the labeled boundaries
and we have used Proposition \ref{ger}. We equip this manifold with the prequantum line bundle
\[
L{(\Sigma,\vec{\varphi})}:=\left.\left(\left.\left(L(\Sigma)\times L(D,{\vec{\varphi}})\right)\right|_{\mu^{-1}_\Sigma(\M_T(D))}\right)\right\slash T(\partial\Sigma_l).
\]
As in the free case, a complex structure on $\Sigma$ determines a polarization of the affine symplectic manifold $\M_T(\Sigma\cup_{\partial\Sigma_l}D)$ -remark that $\M_T(D)$ has a canonical polarization-, and we write $\h_{(\Sigma,\vec{\varphi})}$ for its quantization determined by these data. This Hilbert space carries a positive energy representation of the loop group $T(\partial\Sigma_f)$ associated to the ``free'' boundary components. Next, observe that $\hat{\varphi}\in H^0(\partial\Sigma_l,\hat{A})$ labels an irreducible representation of $T(\partial\Sigma_l)$. In this situation we have:
\begin{theorem}
\label{fact}
 There is a canonical isomorphism 
 \[
 \h_{(\Sigma,\vec{\lambda})}\cong\Hom_{\widetilde{T(\partial\Sigma_l)}}\left(\h_{\vec{\lambda}}, \h_{\Sigma}\right).
 \]
\end{theorem}
\begin{proof}
The proof proceeds in the same way as that of  Theorem \ref{qcr}: 
first one shows in precisely the same way that the Hilbert spaces on both sides carry
an irreducible unitary representation of the same generalized Heisenberg group with equal central character. Next, the right hand side comes equipped with an embedding of the prequantum line bundle $L(\Sigma,\vec{\varphi})$: write $l=(l_\Sigma,l_{D})$, unique up to the $T(\partial\Sigma_l)$-action for an element $[l]\in L(\Sigma,\vec{\varphi})$, where $l_\Sigma\in L(\Sigma)$ and $l_D\in L(D,\vec{\varphi})$. The embedding 
is then given by 
\begin{equation}
\label{ev-in}
ev_l(\psi):=\left< l_\Sigma,\psi(l_{D})\right>,
\end{equation}
with $\psi\in \Hom_{\widetilde{T(S)}}\left(\h_{\vec{\varphi}}, \h_{\Sigma}\right)$
By the same reasoning as above, this is an isometry and independent of the choice of representatives for $[l]\in  L(\Sigma,\vec{\varphi})$. Again Theorem \ref{qbundle} now gives the desired result.
\end{proof}
\begin{remark}
\label{degeneration}
With these results, it is interesting to consider the gluing using the 1-parameter family $A_q,~q\in(0,1)$ of annuli. For two Riemann surfaces $\Sigma_1$ and $\Sigma_2$ as in Theorem \ref{qcr}, this results in the family of surfaces $\Sigma_q:=\Sigma_1\cup_{A_q}\Sigma_2$. As remarked in Example \ref{ex-quant} $ii)$, the vacuum vector $\Omega_{A_q}\in\h_{A_q}$ is given by the trace class operator $q^D$.
 We therefore see that whereas the left hand side of Theorem \ref{qcr} in the 
limit $q\to 1$ is simply given by the ``classical gluing'' $\Sigma_1\cup_S\Sigma_2$ as in Proposition \ref{ger}, the right hand side diverges: in this limit, the operator $q^{D}$ converges to the identity operator in each irreducible summand, which is clearly not trace class. On the other hand, the limit $q\to 0$ does not exist for the left hand side, whereas for the right hand side
$
\lim_{q\to 0} q^{D}=\sum_{\lambda\in\hat{A}}\Omega_\lambda\otimes\overline{\Omega}_{\lambda},
$
where $\Omega_{\varphi}\in\h_\varphi$ is the canonical base vector. This follows by the fact that
$D$ is a generator for the energy $\T$-action on $\h_\varphi$, which by definition is positively graded, its only fixed points being the ray defined by the vacuum vector.
\end{remark}
\section{Conformal field theory}
\label{sectioncft}
\subsection{A finite Heisenberg group and its representations}
In this section we will construct a certain finite Heisenberg group whose representation theory 
controls the structure of the conformal field theories of this paper. Its construction is completely 
topological: Let $\Sigma$ be an oriented surface with compact oriented boundaries, and consider the 
homology group $H_1(\Sigma;A)$, which is finite. Denote by $\ell:=[\Lambda^\circ:\Lambda]$ the index of $\Lambda$ inside $\Lambda^\circ$, so that the inner product $\left<~,~\right>$, restricted to $\Lambda$ takes values in $\ell\Z$. The induced 
bilinear form 
\[
\left<~,~\right>:A\times A\rightarrow \Z/\ell\Z,
\]
together with the intersection form on homology determine a $\Z/\ell\Z$-valued antisymmetric form, denoted by $S$. Associated to $S$ is a central extension 
\begin{equation}
\label{fhg}
0\rightarrow\Z/\ell\Z\rightarrow \widetilde{H_1(\Sigma;A)}\rightarrow H_1(\Sigma;A)\rightarrow 0,
\end{equation}
defined as $\widetilde{H_1(\Sigma;A)}=H_1(\Sigma;A)\times\Z/\ell\Z$ with product given by 
\[
(X,m)\cdot (Y,n)=(X+Y,m+n+S(X\cap Y)).
\]
This group is called the finite Heisenberg group associated to the pair $(H_1(\Sigma;A),S)$. The 
reason for this terminology becomes more clear when thinking of $\Z/\ell\Z$ as the group of $\ell$'th 
roots of unity. Notice that both $H_1(\Sigma;A)$ and its Heisenberg extension are finite groups. The 
following proposition classifies and constructs the irreducible representations of this group in which 
the central $\Z/\ell\Z$ acts by roots of unity.
\begin{proposition} 
\label{gfhg}
Irreducible representations of $\widetilde{H_1(\Sigma;A)}$ in which the central $\Z/\ell\Z$ acts via the standard representation are classified and constructed as follows:
\begin{itemize}
\item[$i)$] the irreducible representations are classified by cohomology classes $\vec{\lambda}\in H^1
(\partial\Sigma;\hat{A})$ such that $\delta(\vec{\lambda})=0$.  They are finite dimensional.
\item[$ii)$] Any representation $\h$ naturally decomposes as 
\[
\h=\bigoplus_{{\vec{\lambda}\in H^1(\partial\Sigma;\hat{A})}\atop{\delta(\vec{\lambda})=0}}\h_{\hat{\Sigma}}\otimes\C_{\vec{\lambda}},
\]
where $\h_{\hat{\Sigma}}$ is a unitary representation of $\widetilde{H_1(\hat{\Sigma};A)}$, the 
Heisenberg group associated to the surface $\hat{\Sigma}$, obtained by gluing disks to the 
boundaries of $\Sigma$, and $\delta$ is the connecting homomorphism in the long exact 
cohomology sequence associated to the pair $(\Sigma,\partial\Sigma)$,
\end{itemize}
\end{proposition}
\begin{proof}
Consider the long exact homology sequence associated to the couple $(\Sigma,\partial\Sigma)$: 
$$
\ldots\longrightarrow H_1(\partial\Sigma;A)\stackrel{i_*}{\longrightarrow}H_1(\Sigma;A)\longrightarrow H_1(\Sigma,\partial\Sigma;A)\longrightarrow\ldots
$$ 
The image of $H_1(\partial\Sigma;A)$ is exactly the kernel of the intersection product $S$ and 
therefore the center of the Heisenberg extension \eqref{fhg} equals ${\rm Im}(i_*)\times\Z/\ell\Z$. The 
Pontryagin dual of the first factor coming from $H_1(\partial\Sigma;A)$ is naturally isomorphic to the 
group of elements $\vec{\lambda}\in H^1(\partial\Sigma;\hat{A})$ such that 
$\delta(\vec{\lambda})=0$, where the last condition accounts for the effect that the group acts via the 
homomorphism $H_1(\partial\Sigma;A)\rightarrow H_1(\Sigma;A)$. In view of Theorem \ref{irgh}, 
this gives $i)$. Furthermore, in any isotypical summand of a representation $\h$, we can mod out 
the center to obtain a unitary representation of the nondegenerate Heisenberg group 
$$
\widetilde{H^1(\Sigma;A)}\slash{\rm Im}(i_*)\cong\widetilde{H^1(\hat{\Sigma};A)}.
$$ 
Decomposing 
the representation $\h$ under the action of the center, the result now follows.
\end{proof}
Elements $\vec{\lambda}\in H^1(\partial\Sigma,\hat{A})$ correspond to a ``full labeling'' of the surface $\Sigma$. We divide boundary components of $\Sigma$ into ``incoming'' and ``outgoing'' according to
the induced orientation from $\Sigma$: if it agrees with the one of the parameterization the component
is called outgoing, and otherwise incoming. This splits the labeling set as $\vec{\lambda}=(\vec{\lambda}_{\mbox{\tiny\rm in}},\vec{\lambda}_{\mbox{\tiny\rm out}})$, and with this we 
write 
\[
\h_{(\partial\Sigma,\vec{\lambda})}:=\h^*_{\vec{\lambda}_{\mbox{\tiny\rm in}}}\otimes\h_{\vec{\lambda}_{\mbox{\tiny\rm out}}},
\]
using the fact from  \S \ref{iper} that the elements in $H^1(\partial\Sigma,\hat{A})$ also label the irreducible positive energy representations of $T(\partial\Sigma)$. With this notation, we define
\begin{equation}
\label{def-mf}
\widetilde{E}\left(\Sigma,\vec{\lambda}\right):=\Hom_{\widetilde{T(\partial\Sigma)}}\left(\h_{(\partial\Sigma,\vec{\lambda})},\h_\Sigma\right).
\end{equation}
Of course, these are just the multiplicity spaces of the representation of $\widetilde{T(\partial\Sigma)}$ on $\h_\Sigma$ and with this definition there is a canonical decomposition 
\[
\bigoplus_{\vec{\lambda}\in H^1(\partial\Sigma;\hat{A})} \widetilde{E}\left(\Sigma,
\vec{\varphi}\right)\otimes\h_{(\partial\Sigma,\vec{\lambda})}\stackrel{\cong}{\longrightarrow}\h_\Sigma
\]
Let us now observe the following:
\begin{proposition}
\label{int-rep}
$\widetilde{E}(\Sigma,\vec{\lambda})$ carries an irreducible representation of $\widetilde{H_1(\Sigma;A)}$ with center $H_1(\partial\Sigma,A)$ acting via the character $\vec{\lambda}$. In particular, $\widetilde{E}(\Sigma,\vec{\lambda})$ is finite dimensional.
\end{proposition}
\begin{proof}
By Theorem \ref{fact}, $\widetilde{E}(\Sigma,\vec{\lambda})$ is canonically isomorphic to the 
quantization of $\M_T(\hat{\Sigma})$ equipped with the line bundle $L(\Sigma,\vec{\lambda})$.
The moduli space $\M_T(\hat{\Sigma})$ is finite dimensional affine symplectic and modelled on 
$$
T_{[A]}\M_T(\hat{\Sigma})=H^1(\hat{\Sigma};\torus).
$$ 
Since $\M_T^0(\hat{\Sigma})=H^1(\hat{\Sigma},T)$ is  compact and 
therefore its quantization will be finite dimensional. By Corollary \ref{fgms}, 
$$
\pi_1
(\M_T(\hat{\Sigma}))=H^1(\hat{\Sigma},\Lambda),
$$ 
and thus the quantization will carry an irreducible representation of the Heisenberg extension of $H^1
(\hat{\Sigma};\Lambda^\circ)$, in which the center $H^1(\hat{\Sigma};\Lambda)$ acts trivially. 
Additionally, as is clear from the construction of the line bundle $L{(\Sigma,\vec{\lambda})}$, there is 
a residual action of $H^0(\partial\Sigma;A)$ labeled by the character $\vec{\lambda}\in H^1
(\partial\Sigma;\hat{A})$. By Proposition \ref{gfhg}, this amounts to an irreducible representation of 
the Heisenberg extension of $H_1(\Sigma;A)$. This completes the proof.
\end{proof}
\begin{remark}
The classification theorem \ref{gfhg} involves the condition $\delta(\vec{\lambda})=0$. This can be seen as follows: the argument above proves that $$\tilde{E}\left(\Sigma,\vec{\lambda}\right)\cong \left(\h_{\M_T(\hat{\Sigma}),L(\Sigma,\vec{\lambda})}\right)^{H^0(\hat{\Sigma},T)}.$$ This equals of course the quantization of the component of $\M_T(\hat{\Sigma})$ over which $H^0(\hat{\Sigma},T)$ acts on the restriction of the prequantum line bundle $L(\Sigma,\vec{\lambda})$ by the trivial character. Recall that $\pi_0(\M_T(\hat{\Sigma}))=H_0(\hat{\Sigma};\Lambda)$. Let $D$ be the union of disks used to obtain $\hat{\Sigma}$ from $\Sigma$. The Mayer--Vietoris sequence gives $$\ldots\longrightarrow 
H_0(\partial\Sigma;\Lambda^\circ)\longrightarrow H_0(D;\Lambda^\circ)\oplus H_0(\Sigma;\Lambda^\circ)\longrightarrow H_0(\hat{\Sigma};\Lambda^\circ)\longrightarrow\ldots,$$ which 
determines the character in $H_0(\hat{\Sigma};\Lambda^\circ)$ by which $H^0(\Sigma,T)$ acts: over 
the connected component determined by $\nu\in H_0(\Sigma;\Lambda)$ and $\vec{\mu}\in H_0(D;
\Lambda)$, this is given by the image of $(\nu,\vec{\lambda}+\vec{\mu})\in H_0(D;
\Lambda^\circ)\oplus H_0(\Sigma;\Lambda^\circ)$ under the map above. Taking $H^0
(\hat{\Sigma},T)$-invariants therefore amounts to considering the connected component 
$\M^0_T(\hat{\Sigma})$ with the line bundle $L{(\Sigma,\vec{\lambda})}$ subject to the condition 
$\delta(\vec{\lambda})=0$.
\end{remark}
\subsection{A unitary modular functor}
\label{conformal_blocks}
The aim of this section is to prove that the spaces $\tilde{E}(\Sigma,\vec{\lambda})$ form a 
unitary modular functor over the ``category'' of Riemann surfaces $\e$: this is the category whose 
objects are the natural numbers $\N$ and a morphism in from $m\in\N$ to $n\in\N$ is an isomorphism class of a Riemann surface $\Sigma$ with $m$ incoming and $n$ outgoing, parameterized, boundaries. 
The space of morphisms $\e(m,n)$ forms an infinite dimensional complex manifold with an action
of the group $Diff^+(\partial\Sigma)$ of orientation preserving diffeomorphisms.
Composition of morphisms is given by gluing surfaces, and this defines holomorphic maps
\[
\e(m,n)\times\e(n,k)\to\e(m,k).
\] 
Remark that $\e$ has no identity morphisms and therefore fails to be a category, but other than 
that it is fine, so one can just work with it as if it were a true category. A modular functor
 is in a way an extension of $\e$. We have already seen one example: the determinant line bundles
 $\Det_\Sigma$ form holomorphic line bundles over $\e$, with the crucial gluing isomorphisms \eqref{det}. This is a one dimensional, or central, extension of $\e$. Now we shall use Theorem \ref{qcr}
to define an extension  of $\e$ by a finite dimensional vector bundle, but we must take
care of the fact that the map in that Theorem becomes only canonical after slightly ``thickening'' the boundaries over which one glues. We therefore proceed as follows:
\begin{lemma}
\label{emb-mf}
There exists a canonical embedding
\[
\Phi_{(\Sigma,\vec{\lambda})}:\Det_\Sigma^{\otimes c}\otimes \widetilde{E}\left(\Sigma,\vec{\lambda}\right)\hookrightarrow \h^*_{(\partial\Sigma,\vec{\lambda})}.
\]
\end{lemma}
\begin{proof}
As already remarked, the complex line bundle defined by the ray associated to the unit element
in $\M_T(\Sigma)\hookrightarrow\proj(\h_\Sigma)$ is canonically isomorphic to the determinant 
line bundle $\Det^{\otimes c}_\Sigma$ over $\e(\Sigma)$. Therefore $\alpha\in\Det_\Sigma^{\otimes c}$ determines a unique vector $\Omega_\Sigma^\alpha\in\h_\Sigma$ in this ray. With this, we define 
the map 
\[
\alpha\otimes\psi\mapsto\psi^*\left(\Omega^\alpha_\Sigma\right)\in\h_{\vec{\lambda}_{\mbox{\tiny\rm 
in}}}^*\otimes\h_{\vec{\lambda}_{\mbox{\tiny\rm out}}},
\] 
where $\psi\in \tilde{E}(\Sigma,\vec{\lambda})$. Since this is obviously complex anti-linear, it defines $\Phi_{(\Sigma,\vec{\lambda})}$. It remains to prove that it is an embedding.

For this, we remark that the Hilbert space $\h_\Sigma$ is canonically rigged by
\[
\check{E}_\Sigma\subset\h_\Sigma\subset\hat{E}_\Sigma,
\]
where $\hat{E}_\Sigma=\Gamma_{hol}\left(\M_T(\Sigma),L(\Sigma)\right)$, see \eqref{rigging}.
In particular, the subspace $\check{E}_\Sigma$ forms a dense domain on which the unitary representation $\widetilde{LT}$ extends holomorphically to $\widetilde{LT}_\C$, and this action restricts on the image of $L(\Sigma)\to\check{E}(\Sigma)$ to the defining action as an equivariant line bundle over $\M_T(\Sigma)$. It therefore follows from \eqref{donfactorization} that $\Omega^\alpha_\Sigma\in\check{E}_\Sigma$ is a cyclic vector under the action of $\widetilde{LT}_\C$.
Since $\h_\Sigma$ is discretely reducible, one easily sees that this implies that $\Omega_\Sigma^\alpha$ is also cyclic for the unitary part, $\widetilde{LT}$.
Suppose now that $\psi^*(\Omega_\Sigma^\alpha)=0$. In other words,
\[
0=\left<v,\psi^*\Omega^\alpha_\Sigma\right>_{\h_{\partial\Sigma,\vec{\lambda}}}=
\left<\psi(v),\Omega^\alpha_\Sigma\right>_{\h_{\Sigma}}
\]
for all $v\in \h_{\partial\Sigma,\vec{\lambda}}$. 
For  $v=\tilde{\varphi}^{-1}\cdot\Omega_{\vec{\lambda}}\in\h_{\partial\Sigma,\vec{\lambda}}$  this implies
that 
\[
\left<\psi(\Omega_{\vec{\lambda}}),\tilde{\varphi}\cdot\Omega^\alpha_\Sigma\right>_{\h_{\Sigma}}=0
\]
for all $\tilde{\varphi}\in\widetilde{LT}$. Because $\Omega^\alpha_\Sigma$ is cyclic, this implies that
$\psi=0$, and completes the proof that the map is an embedding. 
\end{proof}
In view of this result, we now define
\begin{equation}
\label{cb}
E\left(\Sigma,\vec{\lambda}\right):=\Det^{\otimes c}_\Sigma\otimes \tilde{E}\left(\Sigma,\vec{\lambda}\right)
\end{equation}
\begin{proposition}
\label{prop-fact}
Let $\Sigma$ be a Riemann surface with parameterized boundaries obtained by gluing two 
surfaces $\Sigma_1$ and $\Sigma_2$ over $S$. There exists a canonical isomorphism
\begin{equation}
\label{factorization}
\bigoplus_{\vec{\lambda}_S\in H^1(S;\hat{A})} E(\Sigma_1,\vec{\lambda}_1,\vec{\lambda}_S)\otimes 
E(\Sigma_2,\vec{\lambda}_S,\vec{\lambda}_2)\stackrel{\cong}{\longrightarrow}E(\Sigma,
\vec{\lambda}),
\end{equation}
where the labeling $\vec{\lambda}=(\vec{\lambda}_1,\vec{\lambda}_2)\in H^1(\partial\Sigma;\hat{A})$ 
is divided into a partial labeling of $\Sigma_1$ and $\Sigma_2$, which is completed by the $\vec{\lambda}_S\in H^1(S;\hat{A})$
\end{proposition}
\begin{proof}
For notational simplicity, we shall not write out the labels $\vec{\lambda}_1$ and $\vec{\lambda}_2$ in the proof.
Consider the one parameter family $A_q,~q\in(0,1)$ of annuli, and denote by $\Sigma_q$ the result of gluing $\Sigma_1$ and $\Sigma_2$ along both ends of $A_q$. It is known, cf.\cite{segal,kriz} that the 
determinant line bundle forms a one-dimensional modular functor, and there is therefore a canonical isomorphism
\[
\Det_{\Sigma_q}\cong \Det_{\Sigma_1}\otimes\Det_{A_q}\otimes\Det_{\Sigma_2}
\cong \Det_{\Sigma_1}\otimes \Det_{\Sigma_2}
\]
since $\Det_{A_q}$ is canonically trivial. We now claim that there is a commutative diagram
\[
\begin{diagram} \node{\textstyle{\bigoplus_{\vec{\lambda}_S} E(\Sigma_1,\vec{\lambda}_S)\otimes 
E(\Sigma_2,\vec{\lambda}_S)}}\arrow{e,t}{F_q}\arrow{s,t}{\Phi_{(\Sigma_1,\vec{\lambda}_S)}\otimes\Phi_{(\Sigma_2,\vec{\lambda}_S)}}\node{E\left(\Sigma_q\right)}\arrow{s,b}{\Phi_{\Sigma}}\\\node{\h_{\partial\Sigma_1,\vec{\lambda}_S}\otimes\h_{\Sigma_2,\vec{\lambda}_S}}\arrow{e,t}{L_q}\node{\h_{\partial\Sigma}}\end{diagram}
\]
Here $F_q$ is induced by the canonical isomorphism of Theorem \ref{qcr} and the isomorphism
of determinant bundles above, and the vertical maps are as in Lemma \ref{emb-mf}. 
Consulting the construction of the isomorphism of Theorem \ref{qcr}, cf.\ equation \eqref{ev}
we therefore find
\[
F_q\left(\psi_1\otimes\psi_2\right)^*(\Omega^{\alpha_\Sigma}_{\Sigma_q})=\tr_{\h_{\vec{\lambda}}}\left(\left(\psi_1^*\Omega^{\alpha_{\Sigma_1}}_{\Sigma_1}\otimes\psi_2^*\Omega^{\alpha_{\Sigma_2}}_{\Sigma_2}\right) q^D\right),
\]
for $\psi_i\in E(\Sigma_i,\vec{\lambda}_S)$, $i=1,2$ and $\alpha_{\Sigma_i}\in\Det_{\Sigma_i}$. This 
equation defines $L_q$ as $\tr_{\h_{\vec{\lambda}}}(\cdots q^D)$. But since the expression in between
the brackets above is a trace class operator on $\h_{\vec{\lambda}}$, the limit $q\to 1$ of $L_q$ exists, and therefore also for $F_q$. By construction, $F_q$ is an intertwiner of representations of the Heisenberg extension of $H^1(\Sigma,A)$. By continuity, the limit of $F_q$ must be an intertwiner as well, and therefore an isomorphism. This completes the proof.
\end{proof}
We summarize this discussion as follows:
\begin{theorem}
\label{mf}
The assignment $(\Sigma,\vec{\lambda})\mapsto E(\Sigma,\vec{\lambda})$ is part of a unitary 
modular functor, i.e., the vector spaces $E(\Sigma,\vec{\lambda})$ form holomorphic 
vector bundles over the moduli space $\e(\Sigma)$ such that: 
\begin{itemize}
\item[$i)$] (Normalization) $\dim E(\C\proj^1)=1$,
\item[$ii)$] (Tensor property) $E(\Sigma_1\sqcup\Sigma_2)=E(\Sigma_1)\otimes E(\Sigma_2)$,
\item[$iii)$] (Factorization) when $\Sigma=\Sigma_1\cup_S\Sigma_2$, we have a canonical 
isomorphism 
\begin{equation}
\label{factorization}
\bigoplus_{\vec{\lambda}_S\in H^1(S;\hat{A})} E(\Sigma_1,\vec{\lambda}_1,\vec{\lambda}_S)\otimes 
E(\Sigma_2,\vec{\lambda}_S,\vec{\lambda}_2)\stackrel{\cong}{\longrightarrow}E(\Sigma,
\vec{\lambda}),
\end{equation}
where the labeling $\vec{\lambda}=(\vec{\lambda}_1,\vec{\lambda}_2)\in H^1(\partial\Sigma;\hat{A})$ 
is divided into a partial labeling of $\Sigma_1$ and $\Sigma_2$, which is completed by the $\vec{\lambda}_S\in H^1(S;\hat{A})$,
\item[$iv)$] (Unitarity) There exists a canonical nondegenerate ``inner product''
\[
\overline{E\left(\Sigma,\vec{\lambda}\right)}\otimes E\left(\Sigma,\vec{\lambda}\right)\to\overline{\Det}^{\otimes c}_\Sigma\otimes\Det^{\otimes c}_\Sigma,
\]
compatible with factorization.
\end{itemize}
\end{theorem}
\begin{proof}
First remark that by the fact that $E(\Sigma,\vec{\lambda})$ carries an irreducible representation of the Heisenberg extension of $H^1(\Sigma;A)$, it follows that it forms a finite rank holomorphic vector bundle. Using the isomorphism between $E$ and $\tilde{E}$, $i)$ follows by cutting $\C\proj^1$ into two disks and using Schur's Lemma. The fact that $\h_{\Sigma_1\sqcup\Sigma_2}\cong\h_{\Sigma_1}\otimes\h_{\Sigma_2}$, implies $ii)$. Of course $iii)$ is the same as Proposition \ref{prop-fact} Finally, $iv)$ follows from the fact that $\tilde{E}(\Sigma,\vec{\lambda})$ comes equipped with a canonical inner product. By definition \eqref{cb}, we can interpret this as a morphism as stated in $iv)$; Theorem \ref{qcr} shows that this is compatible with factorization.
\end{proof}
\begin{remark}
The list of axioms of a unitary modular functor is the same as those stated in \cite{segal}. 
The proof shows that the relation between factorization and unitarity is quite subtle, and is given by the relation between $E$ and $\tilde{E}$: on the one hand, $\tilde{E}$ has good unitarity properties, but 
its factorization isomorphism can only be given up to a scalar, cf.\ Remark \ref{degeneration}. On the other hand, the spaces of conformal blocks $E$ have canonical factorization properties, however 
their unitarity can only be given as an ``inner product with values in the determinant line bundle''.  
\end{remark}
Finally, let us explain the relation between our construction of the modular functor and its usual definition in terms of invariants. As before, each irreducible representation $\h_\lambda$ carries a  rigging that embeds it into $\hat{E}_{\lambda}:=\Gamma_{hol}(\M_T(D),L_{\lambda})$. This defines 
an embedding of $\h_{(\partial\Sigma,\vec{\lambda})}$ into $\hat{E}_{(\partial\Sigma,\vec{\lambda})}$ defined in the same way where now $D$ is a finite disjoint union of disks. Again, the projective representation of $T(\partial\Sigma)$ extends to a holomorphic action of $T_\C(\partial\Sigma)$. It follows from Stokes' theorem
that the cocycle \eqref{cocycle} is trivial over the subgroup $T^\Sigma_\C\subset T_\C(\partial\Sigma)$ which is therefore isotropic, and there exists a canonical lift $T^\Sigma_\C\to \widetilde{  T_\C(\partial\Sigma)}$
\begin{proposition}
The embedding $\Phi_{(\Sigma,\vec{\lambda})}$ of Lemma \ref{emb-mf} defines an isomorphism
\[
E\left(\Sigma,\vec{\lambda}\right)\stackrel{\cong}{\longrightarrow} \hat{E}_{\partial\Sigma,\vec{\lambda}}^{T^\Sigma_\C}.
\]
\end{proposition}
\begin{proof}
First remark that since $\Omega^\alpha_\Sigma$ is $T^\Sigma_\C$-invariant for all $\alpha\in \Det^{\otimes c}_\Sigma$, the image of the map is contained in the subspace of $T^\Sigma_\C$-fixed vectors. By \cite[Prop12.4]{segal}, $T^\Sigma_\C$ is not only an isotropic subgroup of $T_\C(\partial\Sigma)$, but also positive and compatible with the polarization. Therefore the space of $T^\Sigma_\C$-invariants carries an irreducible representation of the commutant $(T^\Sigma_\C)^\perp$ in $T_\C(\partial\Sigma)$. To compute this commutant, first remark that $\pi_0(T^\Sigma_\C)=H^1(\Sigma,\Lambda)$, and
we can decompose $T^\Sigma_\C=H^0(\Sigma,T_\C)\times V^\Sigma_\C\times H^1(\Sigma,\Lambda)$ with $V^\Sigma_\C:=\torus_\C^\Sigma\slash H^0(\Sigma,\torus_\C)\subset V_\C(\partial\Sigma)$ as boundary values of holomorphic maps. With respect to the symplectic form \eqref{cocycle} we find by Cauchy's theorem,
\[
\left(V^\Sigma_\C\right)^\circ\cong\{\alpha\in\Omega^1_{hol}(\Sigma,\torus_\C),~\alpha|_{\partial\Sigma}=\mbox{exact}\}.
\]
and since the interior of $\Sigma$ is a Stein manifold, we have 
\[ 
\left(V^\Sigma_\C\right)^\circ\slash
V^\Sigma_\C\cong H^1(\hat{\Sigma},\torus_\C).
\]
For the full subgroup $T^\Sigma_\C$ we have the exact sequence
\[
0\to H^0(\Sigma,T_\C)\to T^\Sigma_\C\to\Omega^1_{hol}(\Sigma)\to H^1(\Sigma,T_\C),
\]
where the last map sends $\varphi\in T^\Sigma_\C$ to the holomorphic one form $\varphi^{-1}d\varphi$, which has periods in $\Lambda$. With this we now finally have 
\[
\left.\left(T^\Sigma_\C\right)^\perp\right\slash T^\Sigma_\C\cong H^1(\hat{\Sigma};A).
\]
We can even make the action of this Heisenberg group more explicit: denote, as above, by $D$ the union of disks to close $\Sigma$ to a closed Riemann surface $\hat{\Sigma}$.
By the Atiyah--Bott double coset construction cf.\ \cite{ab},
\[
\M_T(\hat{\Sigma})\cong T^\Sigma_\C\backslash T_\C(\partial\Sigma)\slash T^D_\C,
\]
so that
\[
\hat{E}_{\partial\Sigma,\vec{\lambda}}^{T^\Sigma_\C}\cong\Gamma_{hol}\left(\M_T^0(\hat{\Sigma}),L(\Sigma,\vec{\lambda})\right).
\]
As in Remark \ref{jacobian}, $\M^0_T(\hat{\Sigma})\cong H^1(\hat{\Sigma},T)$ as a symplectic manifold, and we see 
from Proposition \ref{polcl} that the complex structure is given by the isomorphism $H^1(\hat{\Sigma},\torus)\cong H^{0,1}(\hat{\Sigma},\torus_\C)$ given by Hodge theory. The space on the right hand side
of course a finite dimensional space of $\Theta$-functions, on which the Heisenberg group $H^1(\hat{\Sigma},A)$ acts in a well-known fashion. It follows that the spaces have the same dimension, and therefore the map must be an isomorphism.
\end{proof}
\begin{remark}
Theorem \ref{fact} defines a canonical embedding $L(\hat{\Sigma},\vec{\varphi})$ into $\tilde{E}(\Sigma,\vec{\varphi})$ and one easily checks using \eqref{ev-in}, that the embedding $\Phi_{(\Sigma,\vec{\lambda})}$ maps $l\in L(\hat{\Sigma})$ to the natural evaluation
map on $\Gamma_{hol}(\M_T^0(\hat{\Sigma}),L(\Sigma))$. It therefore follows that $\Phi$ is an intertwiner of representations of the Heisenberg group of $H^1(\Sigma,A)$. However, it is not an isometry. Therefore, although the preceding argument proves that the space of $T^\Sigma_\C$-invariants in $\hat{E}_{\partial\Sigma,\vec{\lambda}}$ lies in fact in the Hilbert space $\h_{(\partial\Sigma,\vec{\lambda})}$, the representation of the Heisenberg extension associated to $H^1(\Sigma,A)$ is not unitary with respect to the restriction of the inner product defining the Hilbert space structure.
See \cite{bl} for a alternative derivation of this fact, using the associated projectively flat connection
rather than the factorization property.
\end{remark}

\subsection{The abelian WZW-model}
As explained in \cite{segal}, the existence of a conformal field theory follows from the unitarity of the 
associated modular functor. In this case it is called the abelian WZW-model, which describes 
strings moving on the Riemannian manifold $T$. It is an abelian version of the general
WZW-model, which describes strings moving on an arbitrary compact Lie group. 
\begin{definition}[cf.\ \cite{segal}]
\label{cft}
A unitary conformal field theory is given by a smooth projective monoidal $*$-functor $\Psi$ from the 
complex cobordism category $\e$ to the category $Hilb$ of complex Hilbert spaces and trace class 
maps.
\end{definition}
Notice that the ``category'' of Hilbert spaces and trace class operators fails to be a true category for the same reason as $\e$: the identity mapping on an infinite dimensional Hilbert space is not trace class. Below, we will explicitly spell out the details of this definition.

To define a CFT, i.e., a functor $\Psi:\hat{\e}\rightarrow Hilb$, one needs the following data:
\begin{itemize}
\item[$i)$] a Hilbert space $\h_{S^1}$, such that $\Psi(C_n)=\h^{\otimes n}_{S^1}$, where $C_n$ is the $n$-fold disjoint union of $S^1$. In other words, $\Psi$ is monoidal,
\item[$ii)$] a trace class operator $\Psi_{(\Sigma,\alpha)}:\h_{\partial\Sigma_{\mbox{\tiny in}}}\rightarrow \h_{\partial\Sigma_{\mbox{\tiny out}}}$, for each Riemann surface $\Sigma$ and $$\alpha\in L_\Sigma^{(p,q)}:={\rm Det}_\Sigma^{\otimes p}\otimes\overline{\rm Det}_\Sigma^{\otimes q},~p,q\in\C$$ which only depends on the conformal equivalence class of $\Sigma$,
\end{itemize}
subject to the conditions:
\begin{itemize}
\item The equality $$\Psi_{(\Sigma,\alpha)}=\Psi_{(\Sigma_1,\alpha_1)}\circ \Psi_{(\Sigma_2,\alpha_2)},$$
whenever $\Sigma=\Sigma_1\cup_{C_k}\Sigma_2$ and $\alpha$ equals the image of $\alpha_1\otimes\alpha_2$ under the factorization isomorphism $$L_{\Sigma_1}^{(p,q)}\otimes L_{\Sigma_2}^{(p,q)}\stackrel{\cong}{\longrightarrow}L_\Sigma^{(p,q)}.$$ 
\item $\Psi_{(\overline{\Sigma},\overline{\alpha})}=\Psi^*_{(\Sigma,\alpha)}$, i.e., $\Psi$ is a $*$-functor.
\end{itemize}
A conformal field theory for which $q=0$ is called \textit{chiral}. The pair $(p,q)$ is referred to as the \textit{central charge}. Below we will have $p=q$ and we use this terminology for this single element of $\C$.

\medskip

In our case, the nonlinear $\sigma$-model, we use the theory of positive energy representations of $LT$ to define the basic Hilbert space. For a fixed level in $H^4(BT,\Z)$ define the Hilbert space
\begin{equation}
\label{hswzw}
\h_{S^1}=\bigoplus_{\lambda\in\hat{A}}\h_{\lambda}\otimes\h_{\lambda}^*.
\end{equation}
By the monoidal property, this defines the Hilbert space associated to $C_n$ for all $n\in\N$.
Next, we construct a trace class operator $\Psi_{(\Sigma,\alpha)}:\h_{\mbox{\tiny in}}\rightarrow \h_{\mbox{\tiny out}}$, where $\h_{\mbox{\tiny in}},~\h_{\mbox{\tiny out}}$ are appropriate tensor products of $\h_{S^1}$ corresponding to $\partial\Sigma$, and $\Sigma$ is a Riemann surface with incoming and outgoing boundaries and $\alpha\in L_\Sigma^{(c,c)}$ with central charge given by $c=\dim(\torus)$.

For a labeling $\vec{\lambda}=(\vec{\lambda}_{\mbox{\tiny in}},\vec{\lambda}_{\mbox{\tiny out}})$ of the boundaries of $\Sigma$, we have the finite dimensional Hilbert space $E(\Sigma,\vec{\lambda})$, depending only on the conformal equivalence class of $\Sigma$, i.e., its image in $\e$.  Since $E(\Sigma,\vec{\lambda})$ is finite dimensional the inner product defines a vector $$\Psi^{\left<~,~\right>}_{\Sigma,\vec{\lambda}}\in \tilde{E}(\Sigma,\vec{\lambda})\otimes\overline{\tilde{E}(\Sigma,\vec{\lambda})}.$$ By Proposition \ref{emb-mf}, for $\alpha\in L_\Sigma^{(c,c)}$ we therefore find a vector 
\begin{displaymath}
\begin{split}
\Psi^\alpha_{\Sigma,\vec{\lambda}}:=\alpha\otimes \Psi^{\left<~,~\right>}_{\Sigma,\vec{\lambda}}&\in \Det_\Sigma^c\otimes \tilde{E}(\Sigma,\vec{\lambda})\otimes \overline{\Det}_\Sigma^c\otimes\overline{\tilde{E}(\Sigma,\vec{\lambda})}\\ &\subseteq \left(\h^*_{\vec{\lambda}_{\rm \tiny in}}\otimes\h_{\vec{\lambda}_{\rm \tiny out}}\right) \otimes \overline{\left(\h^*_{\vec{\lambda}_{\rm \tiny in}}\otimes \h_{\vec{\lambda}_{\rm\tiny out}}\right)}\\
&\subseteq \Hom\left(\h_{(\partial\Sigma_{\rm \tiny in},\vec{\lambda}_{\rm \tiny in})},\h_{(\partial\Sigma_{\rm \tiny out},\vec{\lambda}_{\rm \tiny out})}\right)
\end{split}
\end{displaymath}
which is clearly trace-class. With this, we put 
\begin{equation}
\label{defwzw}
\Psi_{(\Sigma,\alpha)}:=\bigoplus_{\vec{\lambda}\in H^1(\partial\Sigma;\hat{A})}\Psi^\alpha_{\Sigma,\vec{\lambda}}:\h_{\partial\Sigma_{\mbox{\tiny in}}}\rightarrow\h_{\partial\Sigma_{\mbox{\tiny out}}},
\end{equation}
 a trace class operator between tensor products of the Hilbert space \eqref{hswzw}. We now arrive at the main conclusion:
\begin{theorem}[``Abelian $WZW$-model'']
\label{thm-wzw}
The Hilbert space $\h_{S^1}$ and the operators $\Psi_{(\Sigma,\alpha)}$ constitute a conformal field theory with central charge $c=\dim(\torus)$. 
\end{theorem}
\begin{proof}
Most importantly, we have to check the composition property, i.e., that $\Psi$ indeed defines a functor from a central extension of $\e$ to $Hilb$. By Theorem \ref{mf}, the map $(\Sigma,\vec{\lambda})\mapsto E(\Sigma,\vec{\lambda})$ forms a unitary modular functor, and therefore the canonical morphism
$$\bigoplus_{\vec{\lambda}_S\in H^1(S,\hat{A})} E\left(\Sigma_1,\sqcup\Sigma_2,\vec{\lambda},\vec{\lambda}_S,\vec{\lambda}_S\right)\otimes 
\overline{E\left(\Sigma_1,\sqcup\Sigma_2,\vec{\lambda},\vec{\lambda}_S,\vec{\lambda}_S\right)}\stackrel{\cong}{\longrightarrow}E\left(\Sigma,
\vec{\lambda}\right)\otimes \overline{E\left(\Sigma,\vec{\lambda}\right)}
$$
induced by factorization will map the vector $$\sum_{\vec{\lambda}_S\in H^1(S,\hat{A})} \Psi^{\left<~,~\right>}_{\Sigma_1\sqcup\Sigma_2,\vec{\lambda},\vec{\lambda}_S,\vec{\lambda}_S}\otimes\alpha_{\Sigma_1\sqcup\Sigma_2}$$ to the 
vector $\Psi^{\left<~,~\right>}_{\Sigma,\vec{\lambda}}\otimes\alpha_\Sigma\in E(\Sigma,\vec{\lambda})\otimes \overline{E(\Sigma,\vec{\lambda})}$, where $\alpha_\Sigma\in L^{c,c}_\Sigma$ is the image of $\alpha_{\Sigma_1}\otimes\alpha_{\Sigma_2}$ under the factorization isomorphism \eqref{det}. This proves functoriality of $\Psi$. It is a $*$-functor, $\Psi^*_{(\Sigma,\alpha)}=\Psi_{(\overline{\Sigma},\overline{\alpha})}$ because of property $iii)$ in Theorem \ref{mf}.
\end{proof}
A Riemannian metric $g$ on $\Sigma$ induces a hermitian metric $\left<~,~\right>_g$ on $\Det_\Sigma$ by means of the $\zeta$-determinant of $\bar{\partial}$ as in \cite{quillen}. As for the conformal
blocks, after taking the $c$-fold tensor product, we view this as an element
\[
\alpha_g\in\Det^{\otimes c}_\Sigma\otimes\overline{\Det}_\Sigma^{\otimes c},
\]
that we can use to define the partition function $\Psi_{\Sigma,g}\in\h_{\partial\Sigma}$ using the previous theorem. It is in this sense that a metric on $\Sigma$ is finally needed to define the partition 
function of a CFT.
\begin{remark}
Again, for odd levels, the construction uses a choice of spin structure. One therefore finds a so-called \textit{spin}-conformal field theory. 
In this case, it is the $2$-dimensional part of the theory described in \cite{moore}.
\end{remark}
The preceding theorem does not a priori refer to the representation of the loop $LT$ on $\h_{S^1}$, although of course its construction depends 
heavily on this structure. There is a way to ``gauge'' this symmetry, which leads to the following structure: in Proposition \ref{emb-mf}, we used the 
basepoint of $\M_T(\Sigma)$, i.e., the trivial bundle, and the determinant line $\Det^c_\Sigma$ to construct the embedding which lead to the trace 
class operator of Theorem \ref{thm-wzw}. Instead, we can use any other point in $\M_T(\Sigma)$, i.e., any other holomorphic $T_\C$-bundle $E$, 
and the line $\Det_{\Sigma,E}^c$ associated to $E$ via the embedding of Proposition \ref{emb-mf}. This leads to the following:
\begin{theorem}[``Gauged abelian WZW-model'']
For any holomorphic $T_\C$-bundle $E$ over $\Sigma$ and $\alpha\in \overline{\Det}_{\Sigma,E}^c\otimes\Det_{\Sigma,E}^c$, there is a unique trace class operator $$\Psi^\alpha_{\Sigma,E}:\h_{\partial\Sigma_{\rm\tiny in}}\rightarrow\h_{\partial\Sigma_{\rm \tiny out}},$$ satisfying composition rules when composing complex cobordisms with holomorphic $T_\C$-bundles.
\end{theorem}

\subsection{The category of Positive energy representations of $LT$} Recall the definition \ref{per} of a positive energy representations of $LT$. In the following, we will restrict to representations with the following property: when decomposed into irreducibles, each of the multiplicity spaces is required to be finite dimensional. Let $\mathcal C$ denote the category 
of such representations at a given level, with bounded intertwiners as morphisms. By the assumption above this category is abelian, and clearly semi-simple. Consider the bifunctor $\left<~,~\right>:\cat^{\mbox{\tiny op}}\times\cat \rightarrow\underline{Hilb}$, defined by 
\[
\left<\h_1,\h_2\right>:=\bigoplus_{\varphi\in\hat{A}}\Hom_{\widetilde{LT}_{op}}\left(\h^*_\varphi,,\h^*_1\right)\otimes\Hom_{\widetilde{LT}}\left(\h_\varphi,\h_2\right),
\] 
where $\underline{Hilb}$ is the category of finite dimensional Hilbert spaces. Thinking of $\left<~,~\right>$ as an ``inner product'', it furnishes the category $\cat$ with the structure of a 2-Hilbert space \cite{baez}. Most important for us, any additive functor between 2-Hilbert spaces has an adjoint, unique up to natural isomorphism, determined by the usual formula familiar from the adjoint of an operator on a Hilbert space. 

For a compact $1$-manifold $S$ let $\cat_S$ be the category of positive energy representations of
$T(S)$ with the above property of having finite multiplicity, so that $\cat_{S^1}=\cat$.  For a complex cobordism $\Sigma$, abbreviate $\cat_{\partial\Sigma_{\mbox{\tiny in}}}, \cat_{\partial\Sigma_{\mbox{\tiny out}}}$ by $\cat_{\mbox{\tiny in}}$, $\cat_{\mbox{\tiny out}}$. The fundamental structure theorem for the category $\cat$ is given as follows:
\begin{theorem}
\label{gf}
Any cobordism $\Sigma$ induces a functor $$U_\Sigma:\cat_{\mbox{\tiny in}}\rightarrow\cat_{\mbox{\tiny out}},$$
satisfying the following properties:
\begin{enumerate}
\item[$i)$] when $\Sigma=\Sigma_1\cup_C\Sigma_2$, there is a natural transformation $$U_\Sigma\cong U_{\Sigma_2}\circ U_{\Sigma_1},$$
\item[$ii)$] $U_A\cong id_\cat$ for any $A$ with the topology of a cylinder,
\item[$iii)$] there is a natural isomorphism $$U_\Sigma^*\cong U_{\overline{\Sigma}},$$
\item[$iv)$] the group ${\rm Diff}^+(\Sigma)$ acts on $U_\Sigma$ by natural transformations, and this action factors over the identity component of ${\rm Diff}^+(\Sigma,\partial\Sigma)$,
\item[$v)$] a complex structure on $\Sigma$, together with an element $\alpha\in {\rm Det}_\Sigma^c$, defines a map $\psi_{\Sigma,\alpha}:E\rightarrow U_\Sigma(E),$ for all objects $\h$ of $\cat_{\mbox{\tiny in}}$, such that $$\psi_{\Sigma,\alpha}=\psi_{\Sigma_2,\alpha_2}\circ\psi_{\Sigma_1,\alpha_1},$$ under the natural transformation in $i)$.
\end{enumerate}
\end{theorem}
Notice that the structure outlined in the theorem is close to what is called a ``category valued topological quantum field theory'' in \cite{segaltft}. In the  theorem above, we have ignored all structure coming from 3-dimensional topology. However, the structure above suffices to prove, as outlined in \cite{segaltft}, the following:
\begin{corollary}
The category $\cat$ of positive energy representation of $LT$ at level $\ell$ with finite dimensional multiplicity spaces is a modular tensor category.
\end{corollary}
\subsection{The $\sigma$-model for a rational torus}
\label{sigma}
We close this paper with a proof of the equivalence, as stated in the ``new'' introduction to \cite{segal},
between the abelian WZW-model and the $\sigma$-model of a \textit{rational} torus. Let us first recall the construction of the $\sigma$-model of a Riemannian torus, cf. \cite[\S10]{segal}: as before we denote by $T$ a torus with Lie algebra $\torus$, so that $T=\torus\slash\Lambda$, with $\Lambda=\pi_1(T)$. The Riemannian structure, i.e., the inner product $\left<~,~\right>$ on $\torus$ determines a dual torus $T^*:=\torus\slash\Lambda^\circ$, with $\Lambda^\circ$ the dual lattice \eqref{dual-lattice}. The loop groups $LT$ and $LT^*$ are dual to each other, as one can see from the decomposition  \eqref{declg} and the corresponding isomorphism $LT^*\cong\Lambda^\circ\times V(S^1)\times T^*$: both loop groups contain the factor $V(S^1)$ for which the symplectic form \eqref{cocycle} defines a nondegenerate pairing, and $T$ and $\Lambda^\circ$, resp. $T^*$ and $\Lambda$ are Pontryagin dual to one another. 
If we write this pairing as $\ll ~,~\gg:LT\times LT^*\to\T$, there is a cocycle on $LT\times LT^*$ defined as
\[
\psi\left((\varphi_1,\zeta_1),(\varphi_2,\zeta_2)\right):=\ll\varphi_1,\zeta_2\gg,
\]
where $\varphi_i\in LT$ and $\zeta_i\in LT^*$ for $i=1,2$. The associated Heisenberg group is polarized by the fact that $V(LT\times LT^*)=V(S^1)\times V(S^1)$ and $V(S^1)$ carries the standard polarization. The unique irreducible representation of this Heisenberg group defined by this polarization can therefore be realized on the Hilbert space
\[
\h_{S^1}=L^2(\Lambda\times T)\otimes\h_{V(S^1)}\otimes\h_{V(S^1)}^*.
\]
Now given a Riemann surface $\Sigma$ with parametrized boundaries, consider the subgroup
$Z_\Sigma\subset T_\C(\partial\Sigma)\times T^*_\C(\partial\Sigma)$ defined as
\[
Z_\Sigma:=\{(\varphi,\zeta)\in T_\C(\Sigma)\times T_\C^*(\Sigma), ~d\varphi\varphi^{-1}=*\sqrt{-1}d\zeta\zeta^{-1}\}.
\]
As proved in \cite[Prop. 10.8]{segal}, $Z_\Sigma$ is a compatible, positive, maximal isotropic subgroup, and there is a canonical choice for a splitting $\chi_\Sigma:Z_\Sigma\to\C^*$ of the central extension over $Z_\Sigma$. This determines a unique ray $L_\Sigma\subset \h_{\partial\Sigma}$ on which $Z_\Sigma$ acts via $\chi_\Sigma$.  It is easy to check that this defines a conformal field theory as in Definition \ref{cft}. Its central charge is given by $(c,c)$. 

This works for any metric $\left<~,~\right>$, but let us now assume that it is \textit{rational}  on $\Lambda\subset\torus$. This implies that the lattice $\Lambda_\circ:=\Lambda\cap\Lambda^\circ$ is of finite index in both $\Lambda$ and $\Lambda^\circ$, and defines a torus $T_\circ:=\torus\slash\Lambda_\circ$ together with a choice of level $q:\Lambda^\circ\times\Lambda^\circ\to\Z$. The central extension of $LT_0$ defined by $q$ has center
\[
A=\left.\left(\frac{1}{2}\Lambda+\frac{1}{2}\Lambda^\circ\right)\right\slash\Lambda_\circ.
\]
The relation between the loop groups of the three tori is given by the exact sequence
\begin{equation}
\label{seq-rat}
0\to A\to LT_\circ\times LT_\circ\to LT\times LT^*\to A\to 0,
\end{equation}
where the first map is the diagonal inclusion of the center, the middle is given by $(\varphi_1,\varphi_2)\mapsto(\varphi_1+\varphi_2,\varphi_1-\varphi_2)$ and the last map 
is the obvious map on the group of components. One observes that the Heisenberg central extension
pulls back to the central extension $\widetilde{LT}_\circ\times\widetilde{LT}_\circ^{op}$ defined by $q$, and since 
the morphism above clearly preserves the polarization class, the Hilbert space $\h_{S^1}$ decomposes as  in \eqref{hswzw}. 
\begin{theorem}
Under this identification of the Hilbert spaces, the $\sigma$-model associated to a rational torus $T$ identifies with that of the abelian WZW-model associated to $T_\circ$.
\end{theorem}
\begin{proof}
First observe that, after complexification, the exact sequence \eqref{seq-rat} restricts to the sequence
\[
0\to H^0(\Sigma;A)\to T^\Sigma_{\circ,\C}\times T_{\circ,\C}^{\overline{\Sigma}}\to Z_\Sigma\to H^1(\Sigma;A)\to 0,
\]
which is also exact. It therefore follows that the ray determined by the $\sigma$-model satisfies
\[
L_\Sigma\subset\bigoplus_{\vec{\lambda}\in H_0(\Sigma;\hat{A})}E\left(\Sigma,\vec{\lambda}\right)\otimes \overline{E\left(\Sigma,\vec{\lambda}\right)}\subset\h_{\partial\Sigma}.
\]
As remarked before, $\vec{\lambda}\in H_0(\Sigma;\hat{A})$ is equivalent to $\vec{\lambda}\in H^1(\partial\Sigma,\hat{A}),~\delta(\vec{\lambda})=0$. This condition on $\vec{\lambda}$ comes from the kernel of the map $T^\Sigma_{\circ,\C}\times T_{\circ,\C}^{\overline{\Sigma}}\to Z_\Sigma$ in the exact sequence above. To further pin down the ray $L_\Sigma$, one considers its cokernel: 
decomposing the Hilbert space under the action of $T^\Sigma_{\circ,\C}\times T_{\circ,\C}^{\overline{\Sigma}}$, each summand $E(\Sigma,\vec{\lambda})\otimes \overline{E(\Sigma,\vec{\lambda})}$ carries a representation of the abelian group $Z_\Sigma\slash (T^\Sigma_{\circ,\C}\times T_{\circ,\C}^{\overline{\Sigma}})=H^1(\Sigma,A)$, which acts on $L_\Sigma$ by the trivial character. 
One verifies that this representation is nothing but the one coming from the diagonal inclusion 
\[
H^1(\Sigma;A)\to\widetilde{H^1(\Sigma;A)}_{op}\times \widetilde{H^1(\Sigma;A)}.
\]
and the irreducible representation of $ \widetilde{H^1(\Sigma;A)}$ on $E(\Sigma,\vec{\lambda})$ of Proposition \ref{int-rep}. But for this diagonal subgroup, the multiplicity of the trivial representation must be one, and the (up to scalar) unique inner product making the representation unitary, must lies in its isotypical summand. But this is exactly the definition \eqref{defwzw} of the partition function of the abelian WZW-model associated to $T_\circ$, so $\Psi_{(\Sigma,\alpha)}\in L_\Sigma$.
\end{proof}
\begin{remark}
The $\sigma$-model of a torus is in general a non-rational CFT. From this point of view, the Theorem
above explains that precisely for rational tori it becomes a rational theory admitting a factorization into
left and right moving parts.
\end{remark}
\appendix
\section{Generalized Heisenberg groups}
\label{heisenberg}
\subsection{Definition}
\label{def-hg}
Following \cite{fms}, we will define a certain class of infinite dimensional groups called \textit{generalized Heisenberg groups}. Let $A$ be an infinite dimensional abelian Lie group such that Lie$(A)$ is a complete locally convex nuclear topological vector space and $\pi_0(A)$ and $\pi_1(A)$ are finitely generated discrete abelian groups. We assume that there is an exact sequence
\begin{equation}
\label{esghg}
0\rightarrow\pi_1(A)\rightarrow {\rm Lie}(A)\stackrel{\exp}{\longrightarrow} A\rightarrow\pi_0(A)\rightarrow 0,
\end{equation}
which makes Lie$(A)$ into a covering space of $A$, i.e., the exponential map is a local diffeomorphism. It follows that any such group can be written (non-canonically) as $A\cong V\times T\times \pi$, where $V$ is a vector space, $T$ a torus and $\pi$ a discrete group. Let $\psi$ be a bi-multiplicative map $\psi:A\times A\rightarrow\T$. In particular, $\psi$ is a group $2$-cocycle and defines  central extension 
$$1
\rightarrow \T\rightarrow \tilde{A}\rightarrow A\rightarrow 0,$$
in which $\tilde{A}=A\times\T$ with multiplication $$(a_1,z_1)\cdot (a_2,z_2)=(a_1a_2,\psi(a_1,a_2)z_1z_2).$$   The isomorphism class of this extension is determined by the commutator map $s:A\times A\rightarrow \T$ defined by $s(a_1,a_2):=\psi(a_1,a_2)\psi(a_2,a_1)^{-1}$, which is skew-multiplicative. We define $Z(A):=\ker(s)$, so that the center of $\tilde{A}$ is given by the induced central extension of $Z(A).$
It was proved in \cite{fms} that any central extension of $A$ is of this kind, i.e., topologically trivial and defined by a group cocycle $\psi$. The opposite of $\tilde{A}$, denoted $\tilde{A}_{op}$, is the central extension associated to the cocycle $\psi^{-1}$.
\begin{definition}
\label{def-gh}
The extension $\tilde{A}$ is called a Heisenberg group when the commutator pairing is nondegenerate, i.e., $Z(A)=\{e\}$. It is called a generalized Heisenberg group when its center $Z(A)$ is a locally compact abelian group.
\end{definition}
\subsection{The linear case: Gaussian measures}
\label{lc}
As a special case, consider a symplectic vector space $(V,\omega)$. The symplectic form $\omega$ defines a cocycle $\psi=\exp \sqrt{-1}\pi\omega$ on the abelian group $V$ and the associated central extension $\tilde{V}$ as above is what is traditionally called the Heisenberg group. Of course this case is well-known, cf.\ \cite{segalcmp}, but we review the theory for completeness. To construct irreducible representations of this Heisenberg group one needs an extra piece of structure on $V$, called a polarization:
\begin{definition}
\label{pol-vs}
A \textit{polarization} of a symplectic vector space is a compatible positive complex structure. Two complex structures $J_1$ and $J_2$ belong to the same polarization class if and only if $J_1-J_2$ is a Hilbert--Schmidt operator. 
\end{definition}
Of course, a complex structure on $V$ is given by an endomorphism $J:V\rightarrow V$ such that $J^2=-1$. Such a complex structure is called 
\begin{itemize}
\item \textit{compatible} with $\omega$ if $\omega(Jv_1,Jv_2)=\omega(v_1,v_2),~\mbox{for all}~v_1,v_2\in V,$ 
\item \textit{positive} if $\omega(Jv,v)>0,~\mbox{for all}~v\neq 0.$
\end{itemize}
A specific choice of complex structure $J$ belonging to a polarization class induces a hermitian pre-inner product 
\begin{equation}
\label{hermmet}
\left<v_1,v_2\right>=\omega(Jv_1,v_2)+\sqrt{-1}\omega(v_1,v_2). 
\end{equation}
The first term defines a real metric on $V$ that we denote by $Q$. Since, by assumption, $V$ is nuclear, $Q$ determines a family $\mu^t_J,~t>0$ of Gaussian measures on the dual $V^*$, uniquely determined by its Fourier transform $$\int_{V^*}e^{\sqrt{-1}\xi(v)}d\mu_J^t(\xi)=e^{-tQ(v,v)/2},$$ for all $v\in V$, cf.\ \cite[chapter IV]{gv}. This measure is not $V$-invariant. Rather, its quasi-invariance under the translation action of $V$ is given by the Cameron--Martin formula for the Radon--Nikodym derivative:
\begin{equation}
\label{cm}
\frac{d\mu_J^t(\xi+v)}{d\mu_J^t(\xi)}=e^{-\left(\frac{1}{2}Q(v,v)+Q(v,\xi)\right)/t},
\end{equation}
for all $v\in V$. Let $L^2_{hol}(V^*,d\mu)$ be the Hilbert space of square integrable holomorphic functions on $V^*$ with respect to the measure $\mu:=\mu^{1}$, with inner product defined by $$\left<f_1,f_2\right>=\int_{V^*}\bar{f}_1(\xi)f_2(\xi)d\mu(\xi).$$  Let $V$ act on $f\in L^2_{hol}(V^*,d\mu)$ by $$(v\cdot f)(\xi):=e^{-(\left<v,\xi\right>/2+\left<v,v\right>/4)}f(\xi+v).$$ A direct computation using the Cameron--Martin formula \eqref{cm} shows that this action is unitary, i.e., $\left<v\cdot f_1,v\cdot f_2\right>=\left<f_1,f_2\right>$ and defines a representation of the Heisenberg group of $V$. This representation is irreducible and denoted by $\h_V$, or perhaps $\h_{V_J}$ if we want to stipulate the dependence on the chosen complex structure $J$.

\subsubsection{Shale's theorem}
\label{ts}
So far, the irreducible representation $\h_V$ of $\tilde{V}$ depends on the specific choice of a compatible, positive complex structure $J$ on $V$. However, $\h_V$ turns out only to depend on the polarization class determined by $J$. Introduce the infinite dimensional ``Siegel upper half space'' $$\mathcal{J}(V)=\left\{\left.\begin{array}{c}\mbox{ Positive compatible}\\\mbox{complex structures $J'$ on $(V,\omega)$}\end{array}\right|~J-J' \mbox{ is Hilbert--Schmidt}\right\}.$$ There are other descriptions of this space: if we choose a $J\in\calJ(V)$ with decomposition $V_\C=W\oplus\overline{W}$, we have $$\mathcal{J}(V)\cong\left\{\mbox{Hilbert schmidt operators}~T:W\rightarrow \bar{W}\left|\begin{array}{l}
i)~ \omega(Tv_1,v_2)=\omega(Tv_2,v_1)\\
ii)~ 1-TT^*>0.
\end{array}\right.\right\}.
$$
Finally, introduce the \textit{restricted symplectic group} $Sp_{res}(V)$ of the polarized symplectic vector space $V$ by $$Sp_{res}(V):=\{A\in Sp(V),~[A,J]~\mbox{is Hilbert--Schmidt}\},$$ for any $J$ belonging to the polarization class. This group acts transitively on $\mathcal{J}(V)$ and one has $$\mathcal{J}(V)=Sp_{res}(V)/U(V_J),$$ where $V_J$ means the complex pre-Hilbert space formed by $J$ by means of the pre-inner product \eqref{hermmet}. We now have:
\begin{theorem}[Shale \cite{shale}]
Two representations $\h_{V_{J_1}}$ and $\h_{V_{J_2}}$ are equivalent if and only if $J_1-J_2$ is Hilbert--schmidt, i.e., if $J_1$ and $J_2$ belong to the same polarization class. 
\end{theorem}
In terms of of Gaussian measures, this can be derived from the following: Let $J_T\in\mathcal{J}(V)$ be another complex structure given in terms of a Hilbert--Schmidt map $T:\bar{A}\rightarrow A$. The Gaussian measure $\mu_{J_T}$ is absolutely continuous with respect to $\mu_J$ with Radon--Nikodym derivative 
\begin{equation}
\label{trgm}
\frac{d\mu_{J_T}(\xi)}{d\mu_J(\xi)}=\det(1-T^*T)^{-1/2}\exp\left(-\frac{1}{2}\left<T\xi,T\xi\right>\right),
\end{equation}
cf.\ \cite[Thm. 4.5]{bsz}. In this formula, $\det(\ldots)$ is the Fredholm determinant of an operator of the form $1+\mbox{trace class}$.

By Schur's lemma, the intertwiner establishing the equivalence of two representations is unique up to a scalar. Consequently, the Hilbert space $\h_{V}$ carries a projective unitary representation of $Sp_{res}(V)$ which extends the representation of $\tilde{V}$ to the semi-direct product $Sp_{res}(V)\ltimes V$. This is the infinite dimensional analogue of the metaplectic representation. 

\subsection{Polarizations and irreducible representations}
Returning to the general case of a generalized Heisenberg group $\tilde{A}$, notice that the commutator pairing induces a skew symmetric bilinear form $$S:\Lie(A)\times\Lie(A)\rightarrow \R,$$ with finite dimensional kernel $\ker(S)\subseteq\Lie(A)$. The quotient vector space $$V(A):=\Lie(A)/\ker(S)$$ is therefore symplectic and in general infinite dimensional. To discuss the representation theory, we need an appropriate generalization of the notion of polarization. Although there seem to be many variants of this, all more or less equivalent, for us the following definition is very useful: 
\begin{definition}[compare {\cite[Def. 10.3]{segal}}]
\label{pol-ghg}
A \textit{polarization} of $A$ means an operator $J:\Lie(A)\rightarrow \Lie(A)$, satisfying
\begin{itemize}
\item[$i)$] $S(J\xi,J\eta)=S(\xi,\eta)$ for all $\xi,\eta\in \Lie(A)$,
\item[$ii)$] $J$ induces a compatible positive complex structure on $V(A)$.
\end{itemize}
Two polarizations define the same \textit{polarization class} if they induce the same polarization class for the symplectic vector space $V(A)$.
\end{definition}
Notice that to define the polarization class on $A$, it suffices to give the operator $J$ up to Hilbert--Schmidt operators. 
The classification theorem of irreducible representations of polarized generalized Heisenberg groups is now as follows:
\begin{theorem}\cite{fms}\label{irgh}
Let $(A,\psi)$ be a generalized Heisenberg group.
A polarization class and a splitting $\chi:Z(A)\rightarrow\T$ of the induced central extension $$1\rightarrow\T\rightarrow \widetilde{Z(A)}\rightarrow Z(A)\rightarrow 1,$$ determine a unique irreducible representation of $\tilde{A}$ in which the center acts according to $\chi$. For a given polarization class, this defines a bijective correspondence between such splittings and irreducible representations.
\end{theorem}
Notice that the theorem implies that up to isomorphism, the irreducible representation $\h_{A,\chi}$, for fixed character $\chi$, only depends on the polarization class, not just the specific polarization. The theorem above unifies Shale's theorem with Mackey's version of the Stone--von Neumann theorem for locally compact abelian groups --satisfying the assumptions stated in \S \ref{def-hg}-- together with the Fourier transform for abelian groups. 
\begin{example}
\label{loopgroup}
A good example is provided by the central extension of the loop group $LT$ at a given level. As shown in \S \ref{iper} this is a generalized Heisenberg group with center $\T\times A$. In this case, there is a natural polarization determined by the canonical action of the circle on itself: This induces an action
on $L\torus$ and on $V(S^1)=L\torus\slash\torus$ it defines a decomposition $V_\C(S^1)=V_+(S^1)\oplus V_-(S^1)$ into complex isotropic subspaces given by the positive, resp. negative Fourier modes. The thus defined complex structure is given by the Hilbert transform. With this, the class of representations for this polarization class of $LT$ are exactly the positive energy representations in the sense of Definition \ref{per}, and the irreducible representations constructed in \S \ref{iper} give an explicit realization of the classification of Theorem \ref{irgh}.
\end{example}

\subsubsection{Isotropic subgroups and the spectral theorem}
\label{induction}
Let $(A,\psi)$ be a generalized Heisenberg group and denote by $\hat{A}$ the dual group of continuous characters. By Definition \ref{def-gh}, a Heisenberg group comes equipped with a map $$e:A\rightarrow\hat{A},$$ defined as $e(a_1)(a_2):=s(a_1,a_2)$, whose kernel equals $Z(A)$. When $Z(A)=\{e\}$, i.e., $(A,\psi)$ is a Heisenberg group, its image is dense in $\hat{A}$.

Let $B\subset A$ be isotropic. This means that one of the following equivalent conditions is satisfied:
\begin{itemize}
\item the extension $\tilde{B}$ induced by $\tilde{A}$ is abelian,
\item $s|_{B\times B}=1$,
\item $B\subseteq B^\perp$, where $B^\perp:=\{a\in A,~s(a,b)=1,~\mbox{for all}~b\in B\}$.
\end{itemize}
When $B$ is isotropic and maximal with respect to any of the above properties, it is said to be Lagrangian.
Because the induced central extension $\tilde{B}$ is abelian, it must be trivial, however not necessarily in a canonical way. A trivialization is determined by a splitting, i.e., a map $\chi:B\rightarrow\T$ satisfying $$\chi(b_1+b_2)=\chi(b_1)\chi(b_2)\psi(b_1,b_2),$$ for all $b_1,b_2\in B$. As such it determines a lifting $B\hookrightarrow \tilde{B}$ by $b\mapsto (b,\chi(b))$. Alternatively, such a splitting can be viewed as a one-dimensional unitary representation of $\tilde{B}$. Via the splitting $\chi$ over $B$, we can restrict unitary representations $\h$ of $\tilde{A}$ to $B$. The spectral theorem
implies that there exists a disintegration 
\[
\h\cong\int_{\hat{B}}^{\oplus}\h_\zeta d\mu^\chi(\zeta)
\]
so that each $b\in B$ acts on $\h_\zeta,~\zeta\in\hat{B}$ via $\chi(b)$.
In this representation,  $U(a)$ acts on $\h$ with respect to the disintegration as 
\[
(U(a)f)(\zeta) =\sqrt{\frac{d\mu(\zeta+e(a))}{d\mu(\zeta)}}U_\zeta(a)(f(\zeta)),
\] 
where $U_\zeta(a):\h_\zeta\rightarrow \h_{\zeta+e(a)},~\zeta\in\hat{B}$ is a measurable family of unitary operators.  In the formula above, the expression in the brackets is the Radon--Nikodym derivative of the $e(a)$-translate of the measure $\mu$ with respect to itself. Restricted to $B^\perp\subseteq A$, this translation action on $\hat{B}$ is trivial, so we see that $\h_\zeta$ carries a unitary representation of $\tilde{B}^\perp$ for each $\xi\in \hat{B}$. When $\h$ is irreducible, this representation 
must be irreducible almost everywhere on $\hat{B}$.

\bibliographystyle{alpha}
{}
\end{document}